%% file: AmorphousCarbonFracture.tex
\documentclass[aps,prb,amsmath,amssymb,superscriptaddress,reprint]{revtex4-1}

\usepackage{graphicx}
\usepackage{dcolumn}
\usepackage{bm}
\usepackage[mathlines]{lineno}
\usepackage[utf8]{inputenc}
\usepackage[T1]{fontenc}
\usepackage{mathptmx}
\usepackage{import}
\usepackage{color}
\usepackage{natmove}
\usepackage{float}
\usepackage[labelformat=empty, position=top]{subcaption}
\usepackage[export]{adjustbox}

\draft 

\begin{document}

\preprint{AIP}

\title[]{Quantitative prediction of the fracture toughness of amorphous carbon from atomic-scale simulations}

\author{S. Mostafa Khosrownejad}
 \affiliation{Department of Microsystems Engineering, University of Freiburg, 79110 Freiburg, Germany}%
\author{James R. Kermode}%
 \affiliation{Warwick Centre for Predictive Modelling, School of Engineering, University of Warwick, CV4 7AL, Coventry, UK.}%
\author{Lars Pastewka}%
 \altaffiliation[Corresponding author:lars.pastewka@imtek.uni-freiburg.de]{}
 \affiliation{Department of Microsystems Engineering, University of Freiburg, 79110 Freiburg, Germany}%
 \affiliation{Cluster of Excellence livMatS, Freiburg Center for Interactive Materials and Bioinspired Technologies, University of Freiburg, 79110 Freiburg, Germany}

\date{\today}

\begin{abstract}
Fracture is the ultimate source of failure of amorphous carbon (a-C) films, however it is challenging to measure fracture properties of a-C from nano-indentation tests and results of reported experiments are not consistent. Here, we use atomic-scale simulations to make quantitative and mechanistic predictions on fracture of a-C. Systematic large-scale K-field controlled atomic-scale simulations of crack propagation are performed for a-C samples with densities of $\rho=2.5, \, 3.0 \, \text{ and } 3.5~\text{g/cm}^{3}$ created by liquid quenches for a range of quench rates $\dot{T}_q = 10 - 1000~\text{K/ps}$. The simulations show that the crack propagates by nucleation, growth, and coalescence of voids. Distances of $ \approx 1\, \text{nm}$ between nucleated voids result in a brittle-like fracture toughness. We use a crack growth criterion proposed by Drugan, Rice \& Sham to estimate steady-state fracture toughness based on our short crack-length fracture simulations. Fracture toughness values of $2.4-6.0\,\text{MPa}\sqrt{\text{m}}$ for initiation and $3-10\,\text{MPa}\sqrt{\text{m}}$ for the steady-state crack growth are within the experimentally reported range. These findings demonstrate that atomic-scale simulations can provide quantitatively predictive results even for fracture of materials with a ductile crack propagation mechanism.
\end{abstract}

\maketitle

\section{\label{sec:introduction}Introduction}
Amorphous carbon (a-C) has many industrial applications, from electrochemical sensors\cite{zeng_diamond-like_2014} to wear resistant coatings\cite{donnet_tribology_2008}.  Mechanical processes, such as plasticity and fracture, play a crucial role in the performance of a-C in these applications.\cite{kunze2014wear}  Fracture toughness is particularly important for the reliability of coatings.  It is often measured using nanoindentation tests.\cite{FU2017107, machanick1998design, LI1998214}  While nanoindentation is easy to perform, it is nontrivial to analyze.\cite{schiffmann2011determination}  The values reported for fracture toughness of a-C scatter between $K_I=4 - 11 \,  \text{MPa}\sqrt{\text{m}}$.\cite{FU2017107, machanick1998design, LI1998214}  Some of the differences can be associated to the difference in density of testing specimen, although this has not been clearly investigated in the literature.  The difference between results highlight the fact that it is not straightforward to measure fracture properties using nanoindentation methods.

An alternative to experiments is the calculation of material properties from atomic-scale simulations that are in principle predictive. However, calculating fracture properties\cite{sinclair_atomistic_1972,bitzek_atomistic_2015,andric_atomistic_2018} has remained difficult for most materials because the process zone at the crack tip, where the material deforms inelastically, is typically much larger than the accessible simulation cell sizes. Typical sizes range from $1~\text{cm}$ for metals to $100~\text{nm}$ for brittle ceramics while molecular calculations are presently limited to size of few $10~\text{nm}$'s. Molecular calculation of fracture properties for fully developed fracture process zone has therefore been limited to brittle crystals that show little to no plasticity near the crack tip: Silicon\cite{broughton_concurrent_1999,abraham_dynamic_2000,perez_directional_2000,perez_ab_2000,bernstein_lattice_2003,zhu_atomistic_2004,zhu_atomistic_2006,buehler_multiparadigm_2006,buehler_threshold_2007,kermode_low-speed_2008,kermode_macroscopic_2013,gleizer_dissociative_2014,kermode_low_2015} and diamond\cite{pastewka_describing_2008,pastewka_screened_2013}. Exceptions are molecular dynamics (MD) calculations on model Lennard-Jones glasses that can yield insights, such as cavitation in front of the crack tip, but are difficult to translate to real-world materials,\cite{falk_molecular-dynamics_1999,falk_simulation_2000} and a few cases of realistic materials such as amorphous Li-Si.\cite{Khosrownejad_2017}

Atomic-scale fracture simulations can be divided into multiple categories.  Simple equiaxial loading simulations can be used to study bond breaking mechanisms.\cite{fyta2006insights} However, since stress distribution is completely different from the real crack tip, this type of simulation does not reveal the mechanism of fracture at crack tips. The thin strip geometry is often employed for the study of dynamic fracture propagation in MD, due to its simplicity, and the fact that the strain energy release rate is constant and independent of crack length.\cite{kermode_low-speed_2008}  However, in thin strip loading, the ``K-field'' near the crack tip has a superimposed uniaxial T-stress which is known to have significant effect on fracture toughness.\cite{lee2018atomistic,sedighiani2011effect} A third solution, which we also employed in the present paper, is to use the solution of linear-elastic fracture mechanics to control the displacement of the outermost atoms of the simulation domain. This approach eliminates the T-stress and can directly provide fracture parameters, given the condition that plasticity is well contained in the simulation zone. This approach goes back to Sinclair\cite{sinclair_atomistic_1972,sinclair1975influence} and was used subsequently to study crystals\cite{perez_directional_2000,perez_ab_2000,pastewka_describing_2008,pastewka_screened_2013} and amorphous materials\cite{Khosrownejad_2017}.  We note that Sinclair\cite{sinclair1975influence} used flexible boundary conditions while Refs.~\onlinecite{perez_directional_2000,perez_ab_2000,pastewka_describing_2008,pastewka_screened_2013,Khosrownejad_2017} employ rigid boundaries (as we do here). Both techniques yield $K_I^c$, but values converge faster with respect to system size for flexible than for rigid boundaries.\cite{buze_numerical-continuation-enhanced_2020}

A systematic study of elasto-plastic behaviour of a-C with different densities was recently performed by Jana \emph{et al.}\cite{Jana_2019,Jana_2020}, showing that the stress sustained during steady-state flow of a-C follows a Drucker-Prager law at room temperature.\cite{drucker_soil_1952}  Furthermore, we have also performed uniaxial tests at $0~\text{K}$ to measure the yield stress of a-C. From these results, we estimate that the process zone radius in mode I fracture for a stress intensity factor $K_I=10 ~\text{MPa}\sqrt{\text{m}}$, which is in the range of experimental results, is less than $22~\text{nm}$. a-C may therefore be in a sweet spot in which near-tip plasticity occurs but the process zone is small enough such that direct atomistic calculations of early stage fracture are possible. Insights obtained from fracture of a-C may well transfer to other network glasses, such as silica or alumina.

We therefore here apply systematic large-scale, K-field controlled MD simulations to study fracture in a-C as a function of its density.  We study initiation and propagation of cracks up to $10$’s of angstroms of crack length under small-scale yielding conditions, that allow for the measurement of fracture resistance curves and initiation fracture toughness $K_I^{c}$.  This simulation needs on the fly calculation of crack-tip position and adjustment of the boundary conditions which is discussed in detail.  Furthermore, based on a criteria introduced by Drugan, Rice \& Sham\cite{DRUGAN1982447}, we calculate steady state fracture toughness $K_I^{ss}$ under pure mode-I fracture loading.  Finally, these calculations enable us to compare our MD results with experimental measurements.

\section{\label{sec:simulation-method}Simulation method}

We use the screened variant of the Tersoff III potential~\cite{tersoff_modeling_1989,pastewka_screened_2013} for all our simulations. This potential was designed to correctly describe bond-breaking processes~\cite{pastewka_describing_2008} which are fundamental to correctly capture plastic deformation and fracture. We perform quasistatic calculations using overdamped MD to simulate fracture in a-C. Quasistatic simulation ensures a clean separation of mechanics from temperature driven relaxation in amorphous systems.  On the contrary, room temperature MD simulation of an amorphous material does not capture the real relaxation in the experimental system due to the vast difference between time scales of simulation and experimental settings.

\subsection{\label{sec:simulation-method-sample} Sample preparation}

We create our samples by quenching a-C from the melt. The melt equilibrates for $5.0~\text{ps}$\cite{mcculloch2000ab} at $6000~\text{K}$, followed by a linear quench down to $0.1~\text{K}$ at three different rates of $10, \, 100,\, 1000~\text{K/ps}$. Lower quench rates are computationally expensive, thus most of our calculations are done with samples prepared with $1000~\text{K/ps}$.  The slower rates of quench is used to investigate quench rate dependence of the results.  Temperature is controlled using the Langevin thermostat during the whole procedure.  Finally, we relax the box and atomistic degrees of freedom using energy minimization method.  The box is allowed to relax anisotropically with energy change tolerance of $10^{-8}~\text{eV}/\text{atom}$ or root mean square force change tolerance of $10^{-8}~\text{eV}\text{\AA}^{-1}/\text{atom}$ and thus the residual stress is completely removed. Liquid quenches are a standard procedure for the creation of a-C samples\cite{mcculloch_ab_2000,de_tomas_graphitization_2016,de_tomas_transferability_2019,Jana_2019}, since direct simulation of a-C thin films growth\cite{jager_molecular-dynamics_2000,belov_calculation_2002,jager_ta-c_2003,moseler_ultrasmoothness_2005,caro_growth_2018} is prohibitively expensive for the sample sizes considered here.

\subsection{\label{sec:simulation-method-properties} Calculation of mechanical properties}

Mechanical properties of a-C have been reported in previous works\cite{Jana_2019,Jana_2020} at room temperature.  However, all of our fracture simulations are performed at zero temperature. Furthermore, it has been shown that the flow stress of a-C depends on the internal pressure of the cell which can be built up during quenching and mechanical loading.\cite{Jana_2020}  Specifically Jana \emph{et al.}\cite{Jana_2020} observed a Drucker-Prager law with a zero-pressure shear flow stress of $41.2~\text{GPa}$ and an internal friction coefficient of $0.39$.  Thus the stress is a function of internal pressure and relaxing the internal pressure during the mechanical loading would decrease the measured stresses.  In the fracture test we perform here, creation of a crack can lead to pressure relaxation during the test.  Thus, to measure the yield stress we apply a uniaxial tension in the $z$ direction and use a fixed periodic boundary in the $y$ direction together with a periodic but relaxed pressure boundary in the $x$ direction. The boundary conditions in the $y$ direction are similar to the fracture test boundary condition that is carried under plane-strain conditions (see Sec.\ref{sec:simulation-method-bc}) and relaxing the pressure in the $x$-direction mimics the relaxation that occurs in the fracture test when the crack opens.

Here, we are using zero temperature uniaxial loading to calculate elasto-plastic material parameters of a-C.  We use a simulation cell of dimensions $l_x=25~\text{\AA}$, $l_y=22~\text{\AA}$ and $l_z=25~\text{\AA}$.  This cell is quenched with the method explained in Sec. \ref{sec:simulation-method-sample} and loaded in $z$ direction by changing the box size in that direction.  The box size is relaxed in $x$ direction using a barostat with time constant of $1~\text{ns}$ to remove the stress in that direction.  The dynamics is damped with a damping constant of $5~\text{fs}$.

\begin{figure*}
  \centering
  \includegraphics[width=0.90\textwidth]{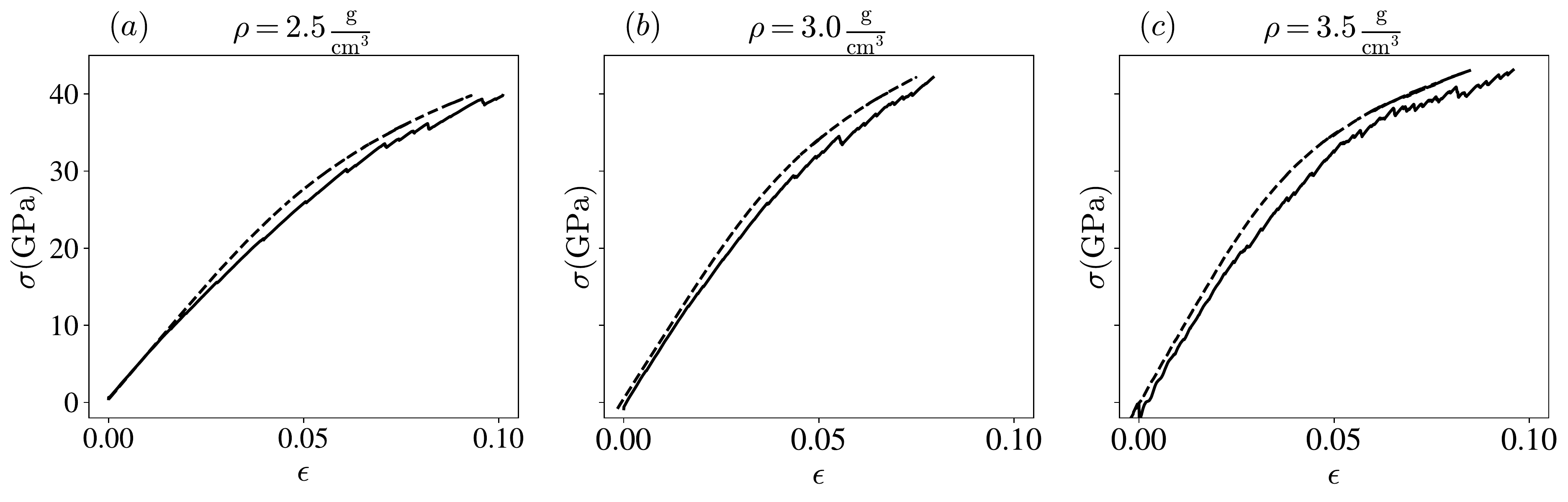} \\
  \caption{Stress strain curve(solid line) and its fitted curve (dashed line) according to the Ramberg–Osgood model for samples with densities $\rho = 2.5, \, 3.0 \text{ and } 3.5 ~\text{g/cm}^{3}$. The stress strain curve is cut at the maximum stress level for each density.}
  \label{fig:yield}
\end{figure*}

We will employ the Hutchinson, Rice \& Rosengren (HRR)\cite{Hutchinson1968,rice_plane_1968} solution of elasto-plastic fracture for the interpretation of our results. HRR uses a Ramberg–Osgood\cite{ramberg_description_1943} plastic material model. We assume that the atomistic samples behave isotropically and $J_2$ plasticity model is a reasonable approximation of their flow behaviour.  Then, the Ramberg-Osgood elasto-plastic material model for a general stress-strain state reads:
\begin{equation}\label{eq:elastoplastic-material}
\varepsilon_{ij} = \frac{1+\nu}{E} s_{ij} + \frac{1-2\nu}{3E} \sigma_{pp} \delta_{ij} + \frac{3}{2E} \alpha \left(\frac{\sigma_e}{\sigma_y}\right)^{n-1} s_{ij}
\end{equation}
Here, $s_{ij}= \sigma_{ij} - \sigma_{pp} \delta_{ij}/3$ is the deviatoric stress, $E$ is the Young's modulus, $\nu$ is Poisson's ratio, $n$ is the hardening power exponent and $\sigma_y$ is the yield strength and we use Einstein summation convention.  The parameter $\alpha$ is associated with the yield strain offset.  In the Ramberg–Osgood model $\alpha \sigma_y/E$ is the strain after which we consider the material to be plastic.  Here we take $\alpha = 0.05$ which is associated with strain deviation of $\approx 0.2\%$ for $\rho=2.5 \, \text{g/cm}^3$.  All other parameters are extracted by fitting the model in Eq.~\eqref{eq:elastoplastic-material} to the simulation results.  Relevant extracted parameters are reflected in Table \ref{tab:mechanical-property}.  The hardening power exponents are first fitted to the data for all samples and calculated to be $n\simeq 5.5$.  To have a consistent plasticity model for all samples we fix the hardening power exponent to be exactly $n=5.5$ for all densities. Fig.~\ref{fig:yield} shows the curve to a sample stress-strain test data.

\begin{table}
\caption{\label{tab:mechanical-property} Mechanical properties of a-C as measured in MD simulation. Estimated plastic zone size is given as a function of $K_I~(\text{MPa}\sqrt{\text{m}})$ and maximum possible $K_I$ is calculated based on plastic zone size estimate.}
\begin{ruledtabular}
\begin{tabular}{ccccccc}
$\rho$& $E$ & $\nu$ & $\sigma_y $ & $r_y$ (\AA) & max $K_I$ & max $K_I$ \\
$\text{g/cm}^{3}$ & $(\text{GPa})$& & $(\text{GPa})$ &$K_I~(\text{MPa}\sqrt{\text{m}})$ & Small& Big sample \\
\hline
$2.5$ & $582$ & $0.216$ & $22.2$ & $2.14 \times K_I^2$ & 4.3 & 10.1\\
$3.0$ & $776$ & $0.218$ & $23.8$ & $1.87 \times K_I^2$ & 4.6 & 10.8\\
$3.5$ & $858$ & $0.229$ & $24.2$ & $1.80 \times K_I^2$ & 4.7 & 11.0
\end{tabular}
\end{ruledtabular}
\end{table}

\subsection{\label{sec:simulation-method-bc} Fracture simulations}
\begin{figure*}
  \centering
  \begin{subfigure}[t]{0.03\textwidth}
    \small (a)
  \end{subfigure}
  \begin{subfigure}[t]{0.42\textwidth}
    \fontsize{10pt}{0pt}\selectfont 
    \def\svgwidth{200pt}
    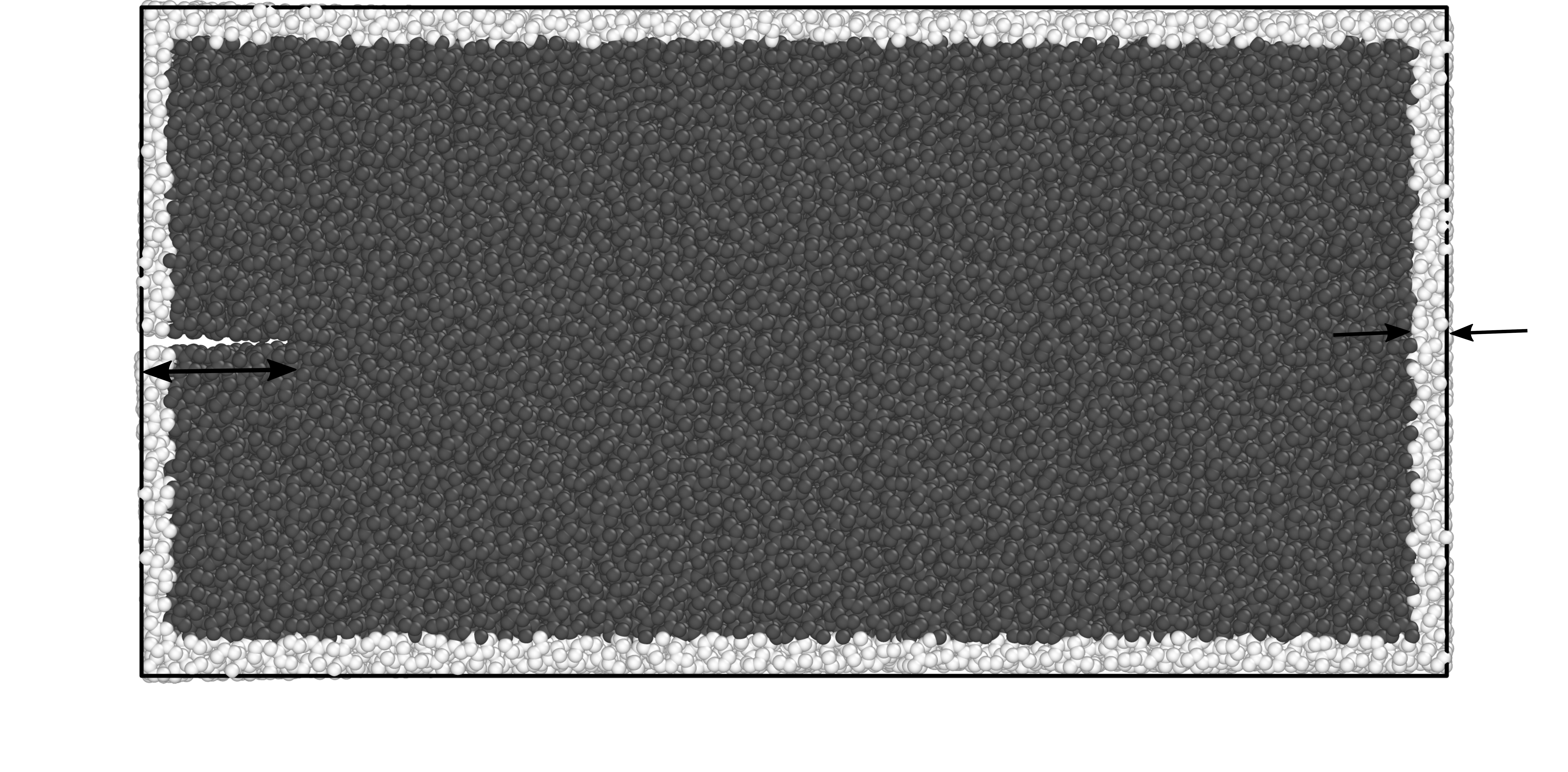
  \end{subfigure}
  \begin{subfigure}[t]{0.03\textwidth}
    \small (b)
  \end{subfigure}
  \begin{subfigure}[t]{0.42\textwidth}
    \includegraphics[width=0.75\textwidth]{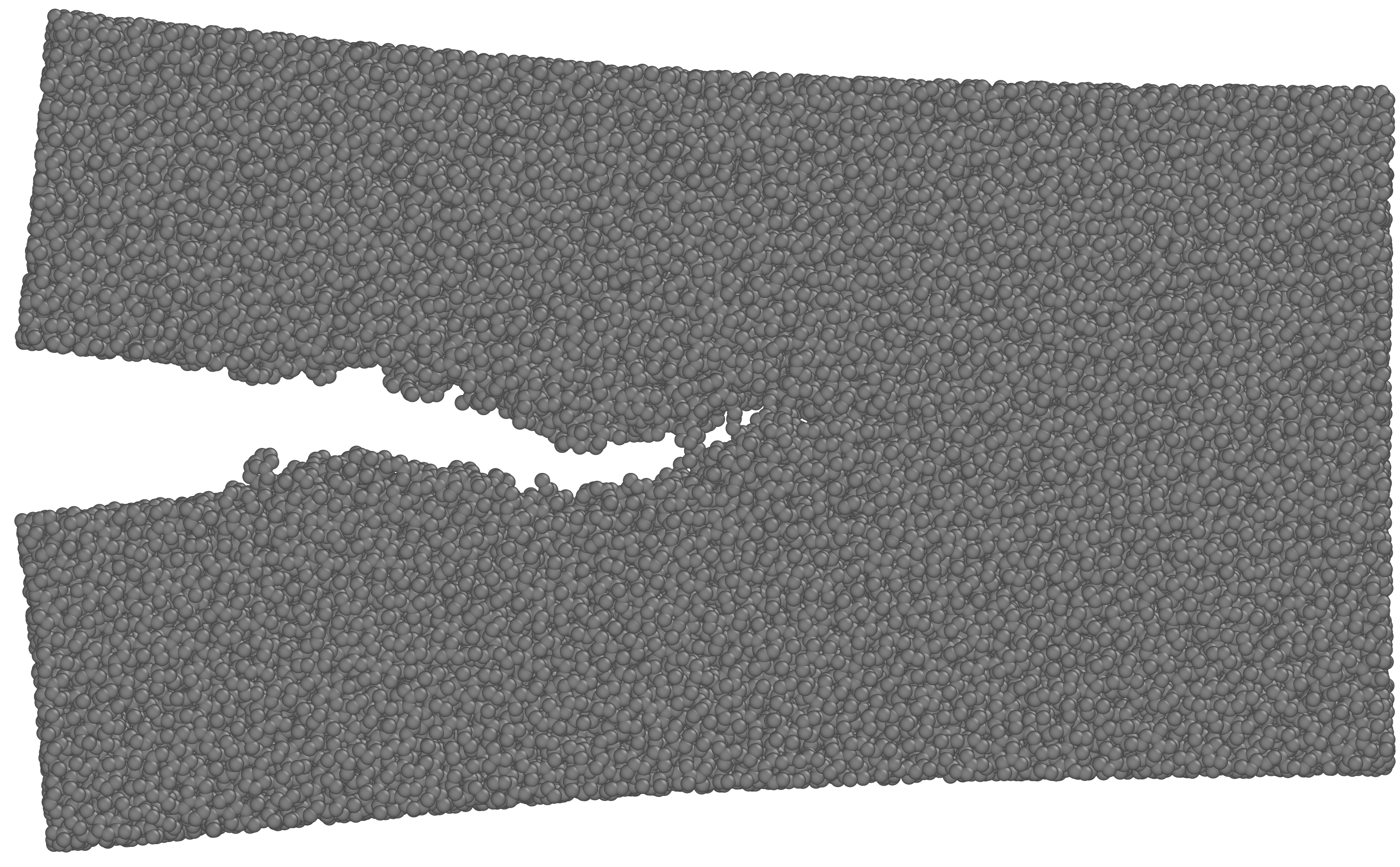}
  \end{subfigure}
  \caption{\label{fig:geometry} (a) Geometry of simulation cell.  Load is applied through K-field displacement of the boundary layer shown with lighter color.  An initial crack of $30~\text{\AA}$ is created through removing bonds between atoms on the opposite sides of the crack surface. (b) Snapshot of fracture in a sample with $\rho=2.5 \, \text{g/cm}^3$ is shown here as a showcase for fracture simulation result.}
\end{figure*}

We study fracture in generalized plane strain conditions.  Plane strain creates smaller process (or plastic) zone than plane stress, thus needs a smaller sample sizes in MD.  We use two different simulation cell sizes.  The simulation cell is limited by the time required for creating the a-C: Quenching bigger samples is prohibitively time consuming for slower quench rates.  The small samples have the dimensions $l_x=100~\text{\AA}$, $l_y=22~\text{\AA}$ and $l_z=200~\text{\AA}$, and big samples have the dimensions $l_x=500~\text{\AA}$, $l_y=22~\text{\AA}$ and $l_z=300~\text{\AA}$.  The cell is periodic in $y$ direction with fixed $l_y$.  We put an initial notch in $x-z$ plane with the length of $30~\text{\AA}$ along the $z$ direction.  Figure~\ref{fig:geometry}a shows the geometry and details of the small simulation cell. An example of a propagated crack configuration if shown in Fig~\ref{fig:geometry}b.

Our molecular calculations follow a technique pioneered by Sinclair\cite{sinclair_atomistic_1972,sinclair1975influence} and subsequently used by many authors.\cite{perez_directional_2000,perez_ab_2000,pastewka_describing_2008,pastewka_screened_2013,Khosrownejad_2017}  We apply displacement boundary conditions on the boundary of the cell compatible with the solution of linear elastic fracture mechanics of isotropic media in mode-I with stress intensity factor $K_I$ for a semi-infinite crack field as follows:
\begin{equation}\label{eq:Kfield}
\begin{array}{lcr}
 \Delta u_x &=& \dfrac{ 2(1+\nu)\Delta K_I}{E} \sqrt{\dfrac{r}{2\pi}}\,\cos(\theta/2)\left[{2-2\nu} - \cos^2(\theta/2) \right]\\
 \Delta u_z &=& \dfrac{ 2(1+\nu)\Delta K_I}{E} \sqrt{\dfrac{r}{2\pi}}\,\sin(\theta/2)\left[{2-2\nu} - \cos^2(\theta/2) \right]
\end{array}
\end{equation}
Here $r$ is the distance from the crack tip and $\theta$ is the angle with respect to the plane of the crack. Loading is applied through incremental increases $\Delta K_I$ of $K_I$, leading to increments of the displacement ($\Delta u_x, \, \Delta u_z$) on the boundary.  The increments of the displacement are applied to the whole sample. Then, the atoms in the boundary region are fixed and the rest are relaxed.  The boundary region has a thickness of $t=10~\text{\AA}$ in the $x-z$ plane. The atoms in the boundary region are fixed relative to each other and therefore they apply a force to other atoms inside the box.  We use increments of $\Delta K_I = 0.05~\text{MPa}\sqrt{\text{m}}$ in each step.  The loading step is followed by a relaxation period of $0.5~\text{ps}$ in NVT ensemble with $T=0.1~\text{K}$ and thermostat damping constant of $5~\text{fs}$.  This is followed by an energy minimization step with an allowed root mean square force change of $10^{-7}~\text{eV}\text{\AA}^{-1}/\text{atom}$ for subsequent minimization steps.

The crack starts moving when $K_I > K_I^c$ and the boundary condition should be adapted accordingly.  We find the new crack tip position after each step of loading using coordination analysis.  Specifically, we detect newly created surface and take the surface atom with the furthest distance along the $z$-direction as the new crack tip position. Namely, we are taking $\max_i z_i$ where $z_i$ are the $z$-positions of all atoms detected by surface detection algorithm as the new crack tip position.

To distinguish the surface from the bulk, a threshold is calculated based on the mean density of the a-C within an atom-centered augmentation sphere of radius $r_c=4~\text{\AA}$. Specifically, given the number of atoms $n_i$ within the sphere centered on atom $i$, the mean density is $\rho_i=m n_i/(4\pi r_c^3/3)$ where $m$ is the mass of a Carbon atom.  When atoms are close to a flat surface, a spherical cap with the height $h$ is removed from the respective sphere in the bulk. This reduces the density from the bulk value $\rho_0$ to $\rho_\text{surf}=(1-\xi^2/4(3-\xi))\rho_0$ with $\xi = h/r_c$. We define $\xi = 0.85$ as the threshold parameter for identification of the surface. All atoms with $\rho_i < \rho_\text{surf}$ are identified as surface atoms.

The important parameter in this surface detection algorithm is the cut-off radius $r_c$.  It should be taken in the same range as the feature sizes that needs to be detected. A too big $r_c$ averages-out the crack-tip features. A small value of $r_c$ introduces noise from the quenched disorder of the a-C's structure into the surface detection algorithm. The value of $r_c=4~\text{\AA}$ and $\xi = 0.85$ are chosen from trial and error. A similar method for crack tip tracking has been used in previous MD simulation of fracture in amorphous Li-Si.\cite{Khosrownejad_2017}

The above method assumes that linear-elastic fracture mechanics is valid some distance away from the crack tip. Analytical and numerical results on elasto-plastic fracture show that this boundary condition remains valid outside plastic zone even for finite deformation elasto-plastic deformation\cite{mcmeeking_finite_1977} and is valid in nanometer proximity to the crack tip for purely brittle materials.\cite{singh_validity_2014}

An estimate for the size of the plastic zone under plane strain conditions based on a continuum mechanical analysis is $r_y=(K_I/\sigma_y)^2/3\pi$\cite{bower2009applied,banks1984form} in the $x$-direction for our geometry. Table~\ref{tab:mechanical-property} shows the anticipated plastic zone size as a function of $K_I$.  The maximum possible $K_I$ load for each density until which the plastic zone can be accommodated is also indicated in Tab.~\ref{tab:mechanical-property} for both sizes of samples.  Applying the boundary condition of Eq.~\eqref{eq:Kfield} is realistic for our simulations, i.e. the plastic zone is confined to the region inside the box until the applied load is less than indicated maximum $K_I$ in Tab.~\ref{tab:mechanical-property}.

An additional correction to the continuum solution (which we apply on the boundary), comes from the quenched disorder of the amorphous atomic structure, leading to stress fluctuations throughout the geometry.  However, Rice\cite{Rice_1974_17} showed that the small deviations from the dominant singular elastic stress, do not have a strong effect on the J-integral or fracture toughness and thus will not alter our simulation results.

\section{\label{sec:simulation-results} Results}
\subsection{\label{sec:r-curve-results} Simulated R-curves}

We used the above-mentioned procedure to measure the crack length $a$, or rather its change $\Delta a$ from the initial condition, as a function of mode-I stress intensity factor $K_I$.  This allows us to plot the function $K_I(\Delta a)$, the crack growth resistance curve often named the R-curve \cite{tvergaard1992relation}.  To avoid confusion, one should bear in mind that we are controlling $K_I$ and measuring $\Delta a$ and the functional notion of R-curve $K_I(\Delta a)$ is used as standard notion in fracture mechanics.  A full R-curve manifests three stages of crack growth: (i) The crack initiation regime provides us with $K_I^c$. (ii) The crack growth regime is characterized by the slope $dK_I/d\Delta a$ that drops toward zero when the crack enters (iii) the steady state regime where the stress intensity factor saturates to a constant fracture toughness value $K_I^{ss}$.  For a semi-infinite crack in a big enough sample loaded according to the elastic K-field, unstable growth only happens if $K_I = K_I^{ss}$.  In an experimental setting unstable crack growth happens when the rate of change of the energy pumped to the crack is greater than the R-curve slope.  This depends on geometry and loading conditions.  Furthermore, geometry and loading of the experimental specimen can change the T-stress (the triaxial stress) which significantly changes the steady state fracture toughness~\cite{Gupta2014review}.

After the crack initiation phase, the crack resistance is a result of energy dissipation in two regions: (i) The fracture process zone, which is a small region close to the crack tip where the fracture processes takes place and, (ii) the plastic zone further away from the crack tip, where plasticity plays the main role and no abrupt change in the micro-structure is visible.  The general understanding in the fracture mechanics community is that the amount of energy dissipated in the fracture process zone remains constant and constitutes the work of separation.\cite{tvergaard1992relation}  This is related to the $K_I^c$ defined above.  For ductile materials, the size of the plastic zone increases with the crack advancement.  The increase in fracture resistance is mainly due to this contribution.  Distinguishing these two effects is the basis for the cohesive zone models of fracture.\cite{tvergaard1992relation}  Due to this separation of effects, it is possible to infer the steady-state fracture toughness $K_I^{ss}$ based on the initiation fracture toughness $K_I^c$, the initial R-curve derivative $dK_I/d\Delta a \big|_{\Delta a =0}$ and the yield stress and we discuss this point in detail in Sec.~\ref{sec:discussion}. Another consequence is that samples with different sizes should lead to the same $K_I^c$. Thus $K_I(\Delta a)$ and simulation results before reaching $K_I^{ss}$ are useful.  We point out that we were unable to reach the steady state in MD simulations due to limitation in computational power which limits the size of the fracture test specimen.  Due to these reasons we only continue the fracture simulation up to $\Delta a \approx 100~\text{\AA}$ and also will be careful in interpreting results of simulations of different sample sizes.

\begin{figure*}
  \centering
  \includegraphics[width=1.0\textwidth]{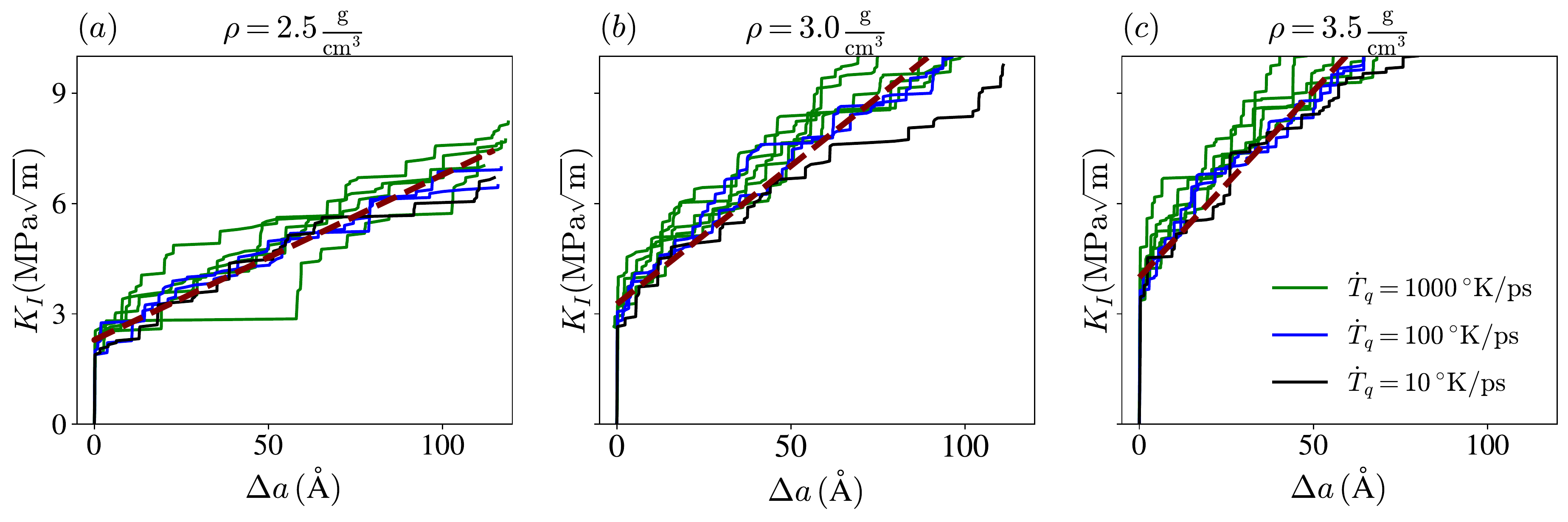} \\
  \caption{R-curves for small samples with densities (a) $\rho=2.5~\text{g/cm}^{3}$, (b) $\rho=3.0~\text{g/cm}^{3}$, (c) $\rho=3.5~\text{g/cm}^{3}$ and three different quench rates $\dot{T}_q$.  Scale of the plots are the same to ease the visual comparison.  The red dashed line is fitted to the samples with quench rate $\dot{T}_q = 1000~\text{K/ps}$.}
  \label{fig:rcurve}
\end{figure*}

We performed simulations with two sample sizes.  The small sample size is chosen in a way to accommodate the plastic zone at the initiation of fracture.  The maximum allowable $K_I$ load is indicated in Tab.~\ref{tab:mechanical-property} in the small sample column.  The smaller size lets us perform the simulation for very slow quench rates.  Figure \ref{fig:rcurve} shows the R-curve of samples with three different densities $\rho = 2.5, \, 3.0 \text{ and } 3.5~\text{g/cm}^{3}$ and three different quench rates $\dot{T}_q = 1000, \, 100, \, 10 ~\text{K/ps}$.  Due to the inherent randomness in the material we provide the result of our test for five different samples for each density and quench rate $\dot{T} = 1000~\text{K/ps}$ that have been melted and quenched independently. The R-curve consists of sections where $K_I$ increases but the crack does not advance and $\Delta a$ is constant, followed by smaller or bigger jumps of the crack tip position. These jumps are accompanied by the breaking of multiple bonds within the system.  The initiation fracture toughness measured in these simulations is smaller than the allowable load for the size of the sample, giving confidence in our methodology.  The R-curve remains linear for higher $K_I$ loads which makes it easy to measure initial R-curve slope $dK_I/d\Delta a\big|_{\Delta a =0}$.  Samples with lower quench rates have the same initiation fracture toughness $K_I^c$, but tend to have smaller R-curve slope.  This effect is however not significant and due to the high computational costs, we do not test slower quench rates for big sample sizes.

\begin{figure*}
  \centering
  \includegraphics[width=1.0\textwidth]{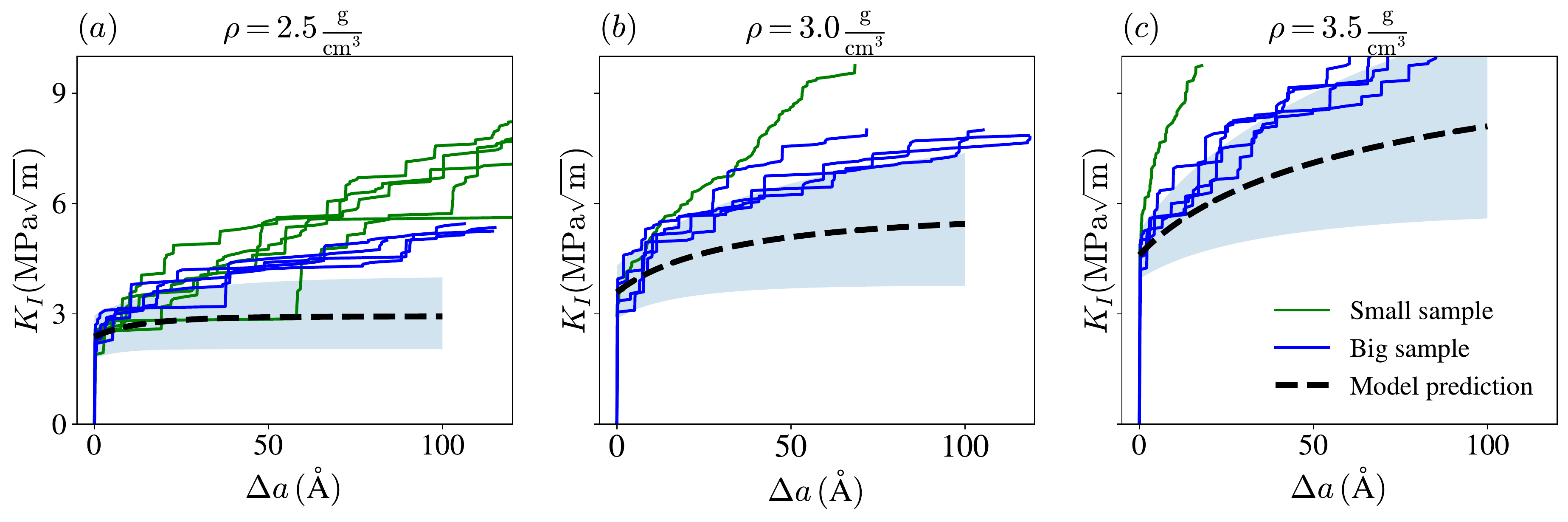} \\
  \caption{R-curves for sample with densities (a) $\rho=2.5~\text{g/cm}^{3}$, (b) $\rho=3.0~\text{g/cm}^{3}$, (c) $\rho=3.5~\text{g/cm}^{3}$, two different sizes and similar quench rates $\dot{T}_q=1000~\text{K/ps}$.  For $\rho=2.5~\text{g/cm}^{3}$ all small sized samples are shown for comparison, but for $\rho = 3.0 \, \text{ and } 3.5~\text{g/cm}^{3}$, only the average values of $\Delta a$ as a function of $K_I$ is shown.  The model prediction is calculated according to Eq. \eqref{eq:k-ode}.  The shaded area shows the 95\% confidence interval.}
  \label{fig:rcurve-big}
\end{figure*}

The big sample can accommodate plastic zone of larger $K_I$ loads which are indicated in Tab.~\ref{tab:mechanical-property} in the big sample column.  Figure \ref{fig:rcurve-big} shows the R-curve of samples with three different densities $\rho = 2.5, \, 3.0, \text{ and } 3.5~\text{g/cm}^{3}$.  For $\rho=2.5~\text{g/cm}^{3}$, we show all small sample and big sample results in one figure.  As can be seen from R-curves for $\rho=2.5~\text{g/cm}^{3}$, initial fracture toughness $K_I^c$ does not statistically change across all samples.  This is in agreement with the separation of effects discussed above.  Furthermore, the initial R-curve slope $dK_I/d\Delta a\big|_{\Delta a =0}$ is also the same for small and big samples.  Small sample and big sample R-curves deviate after the maximum possible $K_I$ load for smaller samples.  For visual clarity, we only show the average value of $\Delta a$ for each $K_I$ for smaller samples and all R-curves for the bigger samples for $\rho = 3.0 \text{ and } 3.5~\text{g/cm}^{3}$.

Table \ref{tab:fracture-property} provides $K_I^c$ and an estimate of $dK_I/d\Delta a\big|_{\Delta a =0}$.  The $K_I^c$ has been calculated based on the R-curves of the big samples.  Due to similarity of the initial R-curve slopes for small and big samples we fit a line to the small sample R-curves with quench rate $\dot{T}_q = 1000~\text{K/ps}$ and report that as an estimate of $dK_I/d\Delta a\big|_{\Delta a =0}$.  This linear regression is shown in Fig.~\ref{fig:rcurve} as a red dashed line. The R-curve slope decreases for the big samples and seems to saturate eventually to a steady state $K_I^{ss}$.  Both initial fracture toughness $K_I^c$ and the initial fracture toughness rate of change $dK_I/d\Delta a\big|_{\Delta a =0}$ increases with the density of sample.  Despite variations between different samples with the same density, the increasing trend in $K_I^c$ and $dK_I/d\Delta a\big|_{\Delta a =0}$ is systematically present.

\begin{table}
\caption{\label{tab:fracture-property} Fracture properties of amorphous carbon as measured in MD simulation.}
\begin{ruledtabular}
\begin{tabular}{ccc}
$\rho~$($\text{g/cm}^{3}$)& $K_I^c$ (MPa$\sqrt{\text{m}}$) & $\frac{dK_I}{d \Delta a}\Big|_{\Delta a = 0}$ (MPa$\sqrt{\text{m}}$/ \mbox{\normalfont\AA}) \\
\hline
$2.5$ & $2.37 \pm 0.29$ & $(0.46 \pm 0.04)\times 10^{-1}$\\
$3.0$ & $3.59 \pm 0.36$ & $(0.75 \pm 0.09)\times 10^{-1}$\\
$3.5$ & $4.60 \pm 0.33$ & $(1.01 \pm 0.21)\times 10^{-1}$
\end{tabular}
\end{ruledtabular}
\end{table}

\subsection{\label{sec:fracture-mechanism} Fracture mechanism}
Given the relatively small experimental fracture toughness values for a-C one might expect a brittle fracture mechanism for the material.  However, this material can also show a relatively large plastic strain\cite{Jana_2019} that creates a a condition for the material to undergo a ductile fracture.  One important advantage of MD simulation of fracture for a-C is that it enables us to better understand the mechanism of fracture and address questions such as this brittle vs ductile fracture morphology.

\begin{figure*}
  \centering
  \begin{subfigure}[t]{0.03\textwidth}
    \small (a)
  \end{subfigure}
  \begin{subfigure}[t]{0.29\textwidth}
    \includegraphics[width=\linewidth, valign=t]{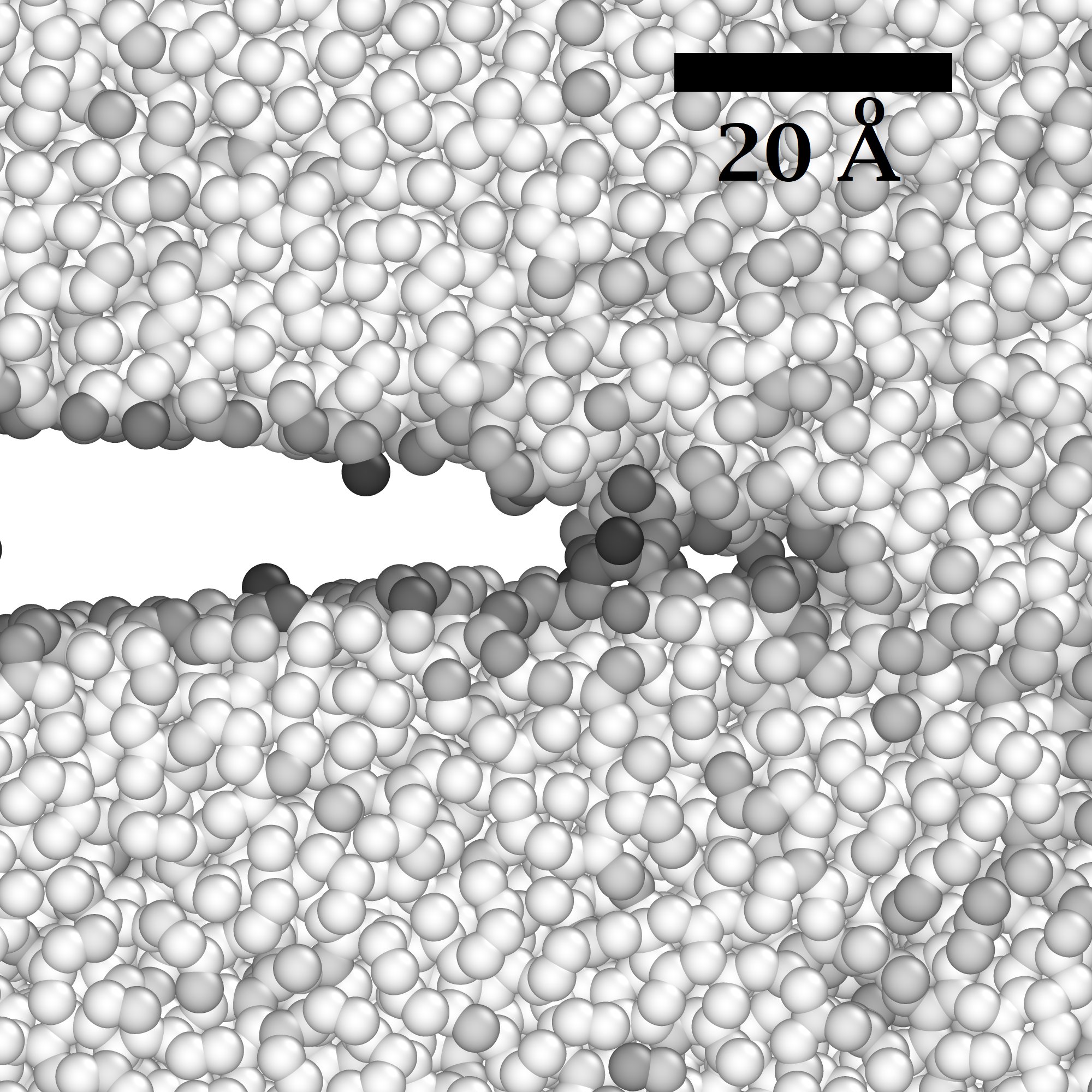}
  \end{subfigure}
  \begin{subfigure}[t]{0.03\textwidth}
    \small (b)
  \end{subfigure}
  \begin{subfigure}[t]{0.29\textwidth}
    \includegraphics[width=\linewidth, valign=t]{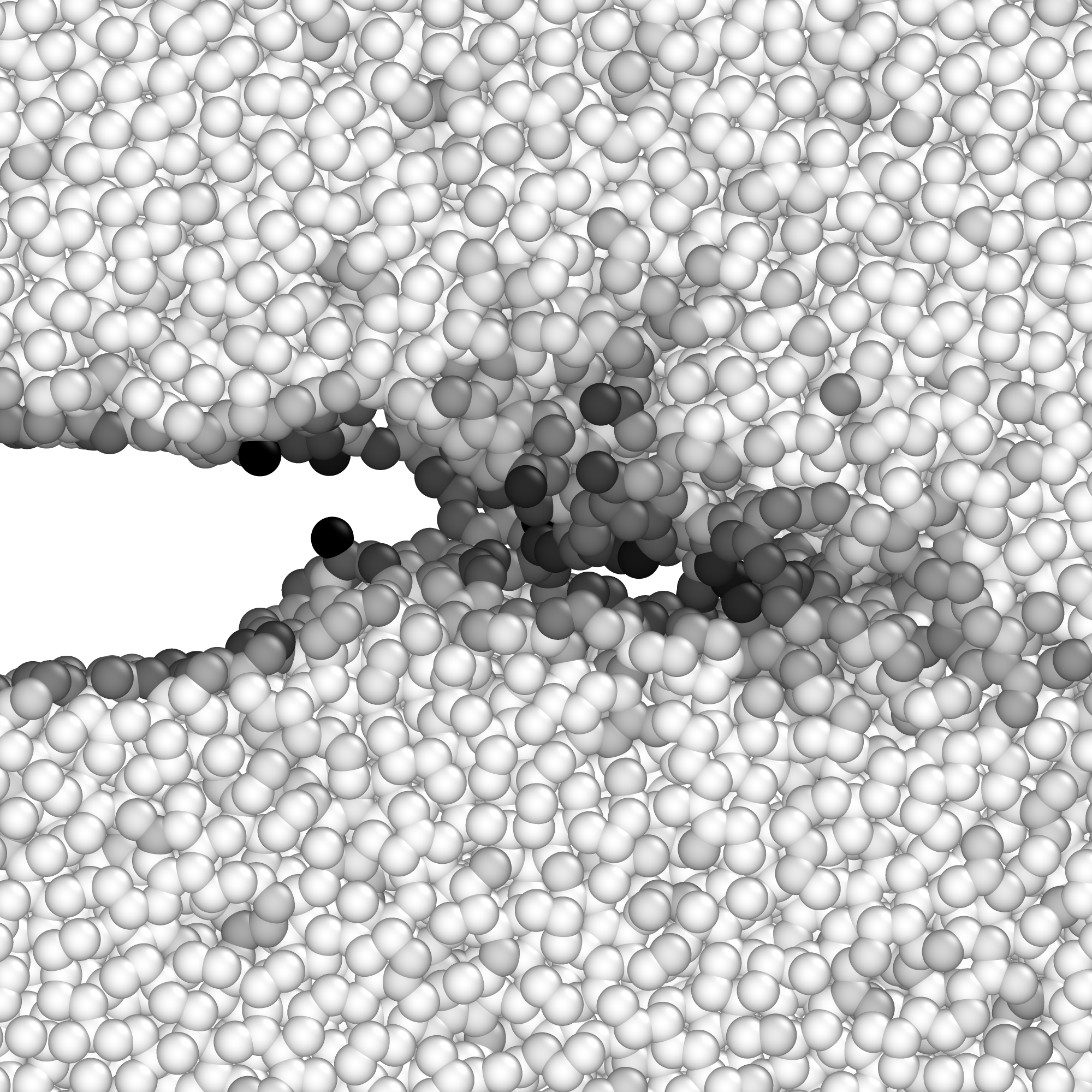}
  \end{subfigure}
  \begin{subfigure}[t]{0.03\textwidth}
    \small (c)
  \end{subfigure}
  \begin{subfigure}[t]{0.29\textwidth}
    \includegraphics[width=\linewidth, valign=t]{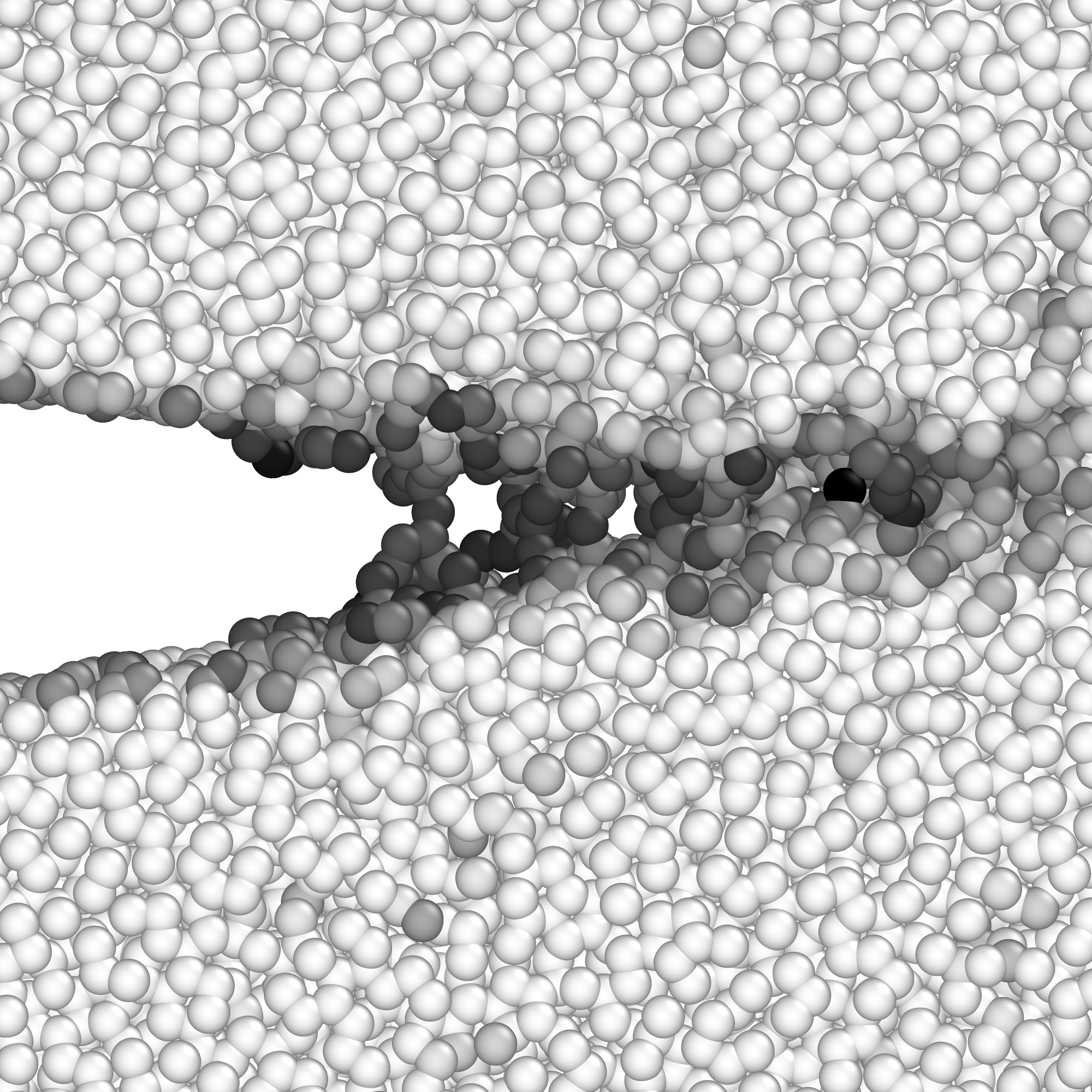}
  \end{subfigure}
  \caption{Snapshots of the creation of voids during fracture process of samples with different densities taken at similar $\Delta a$ (a) $\rho=2.5~\text{g/cm}^{3}$, (b) $\rho=3.0~\text{g/cm}^{3}$, (c) $\rho=3.5~\text{g/cm}^{3}$.  The color code shows the coordination number relative to the average bulk value in each density to better visualize the voids.  Lighter color shows higher coordination number and vice versa. Scale of all snapshots are the same and shown on top of picture a).}
    \label{fig:mechanism}
\end{figure*}

Careful examination of mechanism in MD simulation of a-C shows a consistent mechanism for all densities.  A nano-sized void is formed in front of the crack tip and then crack grows by coalescence mechanism.  Figure \ref{fig:mechanism} shows the void formation for three densities in the course of crack growth.  It is clearly visible that ahead of the crack tip there is a sharply localized zone in which the mean coordination number drops as an indication of void creation.  The small size and distance of voids from the crack tip is responsible for low fracture toughness values and brittle like fracture (see Sec. \ref{sec:discussion}).  The size and the distance of voids from crack tip only slightly changes with density, but generally increases with density.

\section{\label{sec:discussion} Discussion}
\subsection{R-curve trends and qualitative analysis}
Performing careful simulation of mode I fracture loading constitutes a reliable simulation of the fracture process zone.  When we specify an elastic fracture zone with our choice of boundary conditions, a plastic zone emerges inside the sample and a process zone forms near the crack tip.  The experimental value for fracture toughness of a-C is rather low, which is consistent with our simulations.  Thus one expects brittle fracture to take place.  On the contrary, void creation, growth and coalescence mechanism which we report for this material are the typical features of ductile fracture.  According to the Gurson model for ductile fracture, the initiation fracture energy (akin to initiation fracture toughness $K_I^c$) is proportional to void spacing $l^*$ and yield stress, i.e. $J_I^c \propto \sigma_y l^*$\cite{TVERGAARD19921377}.  In our material there are no preexisting voids, but voids are created in the fracture process zone at nanometer-scale distances from the crack tip.  This void creation process suggests a nanoscale value for $l^*$.  Thus, the Gurson model predicts very small values for $K_I^c$ which resembles brittle fracture despite the ductile fracture mechanisms.

Qualitative trends in our R-curves are consistent with the behavior expected from material properties.  From the Gurson model, we expect bigger values for the initiation fracture toughness $K_I^c$ for higher densities: The yield stress and void spacing $l^*$ increase with density and thus according to the Gurson model, we expect an increase in the initiation fracture toughness with increasing density. Furthermore, the increase in $dK_I/d\Delta a\big|_{\Delta a=0}$ for higher densities is consistent with the increase in yield stress and thus higher dissipated energy in plastic zone, and also higher energy needed for tearing of material in the fracture process zone.  The material with higher yield stress and similar micro-mechanism of fracture dissipates more energy in the plastic zone during advancement of the crack.

\subsection{Quantitative predictions and comparison to experiment}

All of our simulations are carried out quasi-statically, essentially corresponding to the $0 ~\text{K}$ limit. There are two competing effects in the mechanical behaviour of amorphous a-C, namely the fast quenching rate which decreases the yield stress and the zero temperature effect which tends to increase the yield stress of a-C.  These competing effects may imply that our simulations on rapidly quenched a-C at zero temperature are more relevant than one might envision.  Nevertheless, we will be careful in correlating fracture and yield stress to extract trends that are related to experimental results.  Furthermore, room temperature simulations would not contribute to a better understanding of the phenomena due to the fact that the competing effects of temperature relaxation and strain rate are occurring at completely different scales in MD simulations and experiments.

There are three reasons that make it difficult to make a direct comparison between MD calculated fracture toughness and experimentally measured values, namely: (i) different geometry of experimental samples which creates complex loading conditions (ii) T-stresses in experimental setting and (iii) different yield stress in MD samples.  In particular, a positive T-stress significantly reduces the fracture energy\cite{XIA1995233}. While very high values of positive residual stress has been reported in deposited a-C thin films\cite{SCHAUFLER2012480}, this is not taken into account for fracture toughness estimates from nano-indentation results.  In general, we expect a deviation from experiment due to T-stresses and geometrical factors.  To be consistent, and rule out the geometrical factors, the two values of $K_I^c$ and $K_I^{ss}$ provide the lower and upper bounds for experimentally measured fracture toughness independent of geometry with zero T-stress.  The values for $K_I^c$ shown in table \ref{tab:fracture-property} for different densities are all in the lower range of experimental values.  Given the fact that the effect of density on the fracture of a-C has not been investigated experimentally, the simulated values provide us with some insight about the effect of density on initiation fracture toughness.

In order to find the steady state fracture toughness, we use a criterion introduced by Drugan, Rice \& Sham\cite{DRUGAN1982447} for crack propagation which is similar to crack growth criterion based on the attainment of a critical accumulated plastic strain\cite{DRUGAN1982447}.  Using this criterion and small scale yielding assumption the crack growth follows a nonlinear differential equation\cite{Ritchie1985,DRUGAN1982447}:
\begin{equation}\label{eq:k-ode}
      T = T_0 - \frac{\beta}{\alpha} \text{ln}\left( \frac{{K_I}^2}{{K_I^c}^2}\right)
       \quad , \quad
       T = \frac{2(1-\nu^2) K_I(\Delta a)}{\sigma_y^2}\frac{dK_I}{d\Delta a}.
\end{equation}
Here $T$ is defined as tearing modulus and $T_0$ is the initial tearing modulus $T_0 = T\big|_{\Delta a = 0}$.  Values of $\alpha$ and $\beta$ depend on the material properties, but $\alpha / \beta \approx 0.1$ is a good approximation for $\nu \approx 0.25$~\cite{Ritchie1985}.  We can find the steady state fracture toughness, using the limit of Eq. \eqref{eq:k-ode} when $\Delta a$ approaches infinity, $K_I \rightarrow K_I^{ss}$ and thus $T \rightarrow 0$:\cite{Ritchie1985}
\begin{equation}\label{eq:kss}
     \frac{K_I^{ss}}{K_I^c}  = \sqrt{\exp\left(\frac{\alpha}{\beta} \, T_0\right)}
\end{equation}
Values of $K_I^c$ and ${dK_I}/{d\Delta a}\Big|_{\Delta a = 0}$ are obtained from our simulations and are given in Tab.~\ref{tab:fracture-property}.  Thus, the criterion enables us to extrapolate to the steady state fracture toughness $K_I^{ss}$ from short crack length MD simulations.  These extrapolated values are presented in Tab.~\ref{tab:estimated-kss}.  Furthermore, the full solution of the differential Eq.~\eqref{eq:k-ode} using the average $K_I^c$ and ${dK_I}/{d\Delta a}\Big|_{\Delta a = 0}$ is shown in Fig. \ref{fig:rcurve-big}.  In Fig.~\ref{fig:rcurve-big}, we also show the upper and lower bounds of the model as shaded area.  The shaded area shows the 95\% confidence interval using two standard deviation $\pm 2\sigma$ difference in $K_I^c$ and ${dK_I}/{d\Delta a}\Big|_{\Delta a = 0}$ (reflected in Tab. \ref{tab:fracture-property}).

\begin{table}
\caption{\label{tab:estimated-kss} Estimation of fracture properties of a-C based on MD simulation results.}
\begin{ruledtabular}
\begin{tabular}{ccc}
$\rho(~\text{g/cm}^{3})$& $K_I^{c}$(MPa$\sqrt{\text{m}}$) & $K_I^{ss}$(MPa$\sqrt{\text{m}}$) \\
\hline
$2.5$ & $2.37 \pm 0.29$ & $2.93^{+0.51}_{-0.46}$ \\
$3.0$ & $3.59 \pm 0.36$ & $5.65^{+1.27}_{-1.03}$ \\
$3.5$ & $4.60 \pm 0.33$ & $9.76^{+3.23}_{-2.30}$
\end{tabular}
\end{ruledtabular}
\end{table}

The values estimated in Table~\ref{tab:estimated-kss} for $K_I^{ss}$ capture the range of reported experimental values in the literature very well, i.e. $K_I = 4-11~\text{MPa}\sqrt{\text{m}}$ \cite{FU2017107, machanick1998design, LI1998214}.  This agreement suggests that the fracture of a-C is very localized and the MD cell sizes used here can capture all the important effects during crack growth.

Table~\ref{tab:estimated-kss} shows that the extrapolated values for $K_I^{ss}$ are lower than the calculated R-curves in Fig.~\ref{fig:rcurve}.  This may be related to the model assumptions of Eq.~\eqref{eq:kss} or ambiguities in the measurement of the yield strength for our atomic-scale a-C models. For example, to the best of our knowledge, there is no direct experimental measurement of the yield strength of a-C in the literature. On the other hand, calculation of yield stress in MD is not straight forward (explained in Sec. \ref{sec:simulation-method-properties}). If we take the fixed volume condition for yield stress measurement (similar to Jana \emph{et al.}\cite{Jana_2020}) the predicted values of yield stress increases over our estimates and thus $K_I^{ss}$ would decrease according to Eq.~\eqref{eq:kss}.  In contrast, performing a test by completely relaxing the pressure during the test would lower the yield stress and thus increase the estimated $K_I^{ss}$ according to Eq.~\eqref{eq:kss}.  Here we choose a trade-off between these two extreme cases by fixing the $l_y$ and relaxing $l_x$ which we believe is the best predictor of the yield stress in the fracture test.

\section{\label{sec:conclusion} Conclusion}
In this paper we performed large scale MD mode-I fracture simulation of a-C with K-field controlled boundary conditions and a well-confined plastic zone. Since experimental measurements of fracture toughness in a-C as a function of density remain challenging, our simulation can provide valuable insight about the fracture process. In this work we simulated fracture of a-C with three different densities $\rho=2.5, \, 3.0 \text{ and } 3.5~\text{g/cm}^{3}$. The fracture mechanism for all samples is void creation, growth and coalescence.  However, due to the small distance between voids, the fracture toughness remains in the usual range for brittle fracture. Both calculated $K_I^c$ and extrapolated $K_I^{ss}$ are in the range of experimentally measured values which is an indication that the fracture process depends on localized effects near the crack tip that are captured in our atomic-scale simulations.

We want to emphasis that the quantitative predictive power of the atomic-scale simulation results of the fracture of a-C is very encouraging and the methodology and insights are applicable for other glassy materials with similar properties. Other than that, atomic-scale simulations allow controlled testing of various assumptions underlying fracture mechanics, e.g. one could test mixed mode fracture scenarios, or test different fracture path selection criteria such as maximum energy release rate or principle of local symmetry. We also propose more accurate experimental methods, such as micro beam bending experiment that have been performed on hydrogenated a-C~\cite{SCHAUFLER2012480}, to better understand fracture of a-C.

\section*{Acknowledgements}

Atomic-scale simulations were carried out with LAMMPS.\cite{plimpton_fast_1995} Postprocessing and visualization were carried out with OVITO.\cite{stukowski_visualization_2010} We acknowledge support from the European Research Council (ERC-StG-757343) and from the EPSRC (EP/R012474/1 and EP/R043612/1). We are indebted to the Jülich Supercomputing Center for allocation of computing time on JUWELS (grant hfr13).
Postprocessing was carried out on NEMO at the University of Freiburg (DFG grant INST
39/963-1 FUGG).

\providecommand{\noopsort}[1]{}\providecommand{\singleletter}[1]{#1}%

\end{document}

%% file: Resources/inkscape/geometry.pdf_tex
\begingroup%
  \makeatletter%
  \providecommand\color[2][]{%
    \errmessage{(Inkscape) Color is used for the text in Inkscape, but the package 'color.sty' is not loaded}%
    \renewcommand\color[2][]{}%
  }%
  \providecommand\transparent[1]{%
    \errmessage{(Inkscape) Transparency is used (non-zero) for the text in Inkscape, but the package 'transparent.sty' is not loaded}%
    \renewcommand\transparent[1]{}%
  }%
  \providecommand\rotatebox[2]{#2}%
  \newcommand*\fsize{\dimexpr\f@size pt\relax}%
  \newcommand*\lineheight[1]{\fontsize{\fsize}{#1\fsize}\selectfont}%
  \ifx\svgwidth\undefined%
    \setlength{\unitlength}{2428.12319699bp}%
    \ifx\svgscale\undefined%
      \relax%
    \else%
      \setlength{\unitlength}{\unitlength * \real{\svgscale}}%
    \fi%
  \else%
    \setlength{\unitlength}{\svgwidth}%
  \fi%
  \global\let\svgwidth\undefined%
  \global\let\svgscale\undefined%
  \makeatother%
  \begin{picture}(1,0.49238701)%
    \lineheight{1}%
    \setlength\tabcolsep{0pt}%
    \put(0,0){\includegraphics[width=\unitlength,page=1]{geometry.pdf}}%
    \put(0.94222108,0.26372714){\color[rgb]{0,0,0}\makebox(0,0)[lt]{\begin{minipage}{0.04202896\unitlength}\raggedright \end{minipage}}}%
    \put(0.93444011,0.2398992){\color[rgb]{0,0,0}\makebox(0,0)[lt]{\lineheight{1.25}\smash{\begin{tabular}[t]{l}10 \AA\end{tabular}}}}%
    \put(-0.00150579,0.22499024){\color[rgb]{0,0,0}\makebox(0,0)[lt]{\lineheight{1.25}\smash{\begin{tabular}[t]{l}30 \AA\end{tabular}}}}%
    \put(0,0){\includegraphics[width=\unitlength,page=2]{geometry.pdf}}%
    \put(0.14732011,0){\color[rgb]{0,0,0}\makebox(0,0)[lt]{\lineheight{1.25}\smash{\begin{tabular}[t]{l}z\end{tabular}}}}%
    \put(0.04278088,0.07344451){\color[rgb]{0,0,0}\makebox(0,0)[lt]{\lineheight{1.25}\smash{\begin{tabular}[t]{l}x\end{tabular}}}}%
    \put(0.10864498,0.03265416){\color[rgb]{0,0,0}\makebox(0,0)[lt]{\lineheight{1.25}\smash{\begin{tabular}[t]{l}y\end{tabular}}}}%
  \end{picture}%
\endgroup%

%% file: AmorphousCarbonFracture.bbl
\begin{thebibliography}{63}%
\makeatletter
\providecommand \@ifxundefined [1]{%
 \@ifx{#1\undefined}
}%
\providecommand \@ifnum [1]{%
 \ifnum #1\expandafter \@firstoftwo
 \else \expandafter \@secondoftwo
 \fi
}%
\providecommand \@ifx [1]{%
 \ifx #1\expandafter \@firstoftwo
 \else \expandafter \@secondoftwo
 \fi
}%
\providecommand \natexlab [1]{#1}%
\providecommand \enquote  [1]{``#1''}%
\providecommand \bibnamefont  [1]{#1}%
\providecommand \bibfnamefont [1]{#1}%
\providecommand \citenamefont [1]{#1}%
\providecommand \href@noop [0]{\@secondoftwo}%
\providecommand \href [0]{\begingroup \@sanitize@url \@href}%
\providecommand \@href[1]{\@@startlink{#1}\@@href}%
\providecommand \@@href[1]{\endgroup#1\@@endlink}%
\providecommand \@sanitize@url [0]{\catcode `\\12\catcode `\$12\catcode
  `\&12\catcode `\#12\catcode `\^12\catcode `\_12\catcode `\%12\relax}%
\providecommand \@@startlink[1]{}%
\providecommand \@@endlink[0]{}%
\providecommand \url  [0]{\begingroup\@sanitize@url \@url }%
\providecommand \@url [1]{\endgroup\@href {#1}{\urlprefix }}%
\providecommand \urlprefix  [0]{URL }%
\providecommand \Eprint [0]{\href }%
\providecommand \doibase [0]{http://dx.doi.org/}%
\providecommand \selectlanguage [0]{\@gobble}%
\providecommand \bibinfo  [0]{\@secondoftwo}%
\providecommand \bibfield  [0]{\@secondoftwo}%
\providecommand \translation [1]{[#1]}%
\providecommand \BibitemOpen [0]{}%
\providecommand \bibitemStop [0]{}%
\providecommand \bibitemNoStop [0]{.\EOS\space}%
\providecommand \EOS [0]{\spacefactor3000\relax}%
\providecommand \BibitemShut  [1]{\csname bibitem#1\endcsname}%
\let\auto@bib@innerbib\@empty
\bibitem [{\citenamefont {Zeng}\ \emph {et~al.}(2014)\citenamefont {Zeng},
  \citenamefont {Neto}, \citenamefont {Gracio},\ and\ \citenamefont
  {Fan}}]{zeng_diamond-like_2014}%
  \BibitemOpen
  \bibfield  {author} {\bibinfo {author} {\bibfnamefont {A.}~\bibnamefont
  {Zeng}}, \bibinfo {author} {\bibfnamefont {V.~F.}\ \bibnamefont {Neto}},
  \bibinfo {author} {\bibfnamefont {J.~J.}\ \bibnamefont {Gracio}}, \ and\
  \bibinfo {author} {\bibfnamefont {Q.~H.}\ \bibnamefont {Fan}},\ }\href
  {\doibase 10.1016/j.diamond.2014.01.003} {\bibfield  {journal} {\bibinfo
  {journal} {Diam. Relat. Mater.}\ }\textbf {\bibinfo {volume} {43}},\ \bibinfo
  {pages} {12} (\bibinfo {year} {2014})}\BibitemShut {NoStop}%
\bibitem [{\citenamefont {Donnet}\ and\ \citenamefont
  {Erdemir}(2008)}]{donnet_tribology_2008}%
  \BibitemOpen
  \bibfield  {author} {\bibinfo {author} {\bibfnamefont {C.}~\bibnamefont
  {Donnet}}\ and\ \bibinfo {author} {\bibfnamefont {A.}~\bibnamefont
  {Erdemir}},\ }\href {\doibase 10.1007/978-0-387-49891-1} {\emph {\bibinfo
  {title} {Tribology of Diamond-Like Carbon Films}}},\ edited by\ \bibinfo
  {editor} {\bibfnamefont {C.}~\bibnamefont {Donnet}}\ and\ \bibinfo {editor}
  {\bibfnamefont {A.}~\bibnamefont {Erdemir}}\ (\bibinfo  {publisher} {Springer
  US},\ \bibinfo {address} {Boston},\ \bibinfo {year} {2008})\BibitemShut
  {NoStop}%
\bibitem [{\citenamefont {Kunze}\ \emph {et~al.}(2014)\citenamefont {Kunze},
  \citenamefont {Posselt}, \citenamefont {Gemming}, \citenamefont {Seifert},
  \citenamefont {Konicek}, \citenamefont {Carpick}, \citenamefont {Pastewka},\
  and\ \citenamefont {Moseler}}]{kunze2014wear}%
  \BibitemOpen
  \bibfield  {author} {\bibinfo {author} {\bibfnamefont {T.}~\bibnamefont
  {Kunze}}, \bibinfo {author} {\bibfnamefont {M.}~\bibnamefont {Posselt}},
  \bibinfo {author} {\bibfnamefont {S.}~\bibnamefont {Gemming}}, \bibinfo
  {author} {\bibfnamefont {G.}~\bibnamefont {Seifert}}, \bibinfo {author}
  {\bibfnamefont {A.~R.}\ \bibnamefont {Konicek}}, \bibinfo {author}
  {\bibfnamefont {R.~W.}\ \bibnamefont {Carpick}}, \bibinfo {author}
  {\bibfnamefont {L.}~\bibnamefont {Pastewka}}, \ and\ \bibinfo {author}
  {\bibfnamefont {M.}~\bibnamefont {Moseler}},\ }\href@noop {} {\bibfield
  {journal} {\bibinfo  {journal} {Tribol. Lett.}\ }\textbf {\bibinfo {volume}
  {53}},\ \bibinfo {pages} {119} (\bibinfo {year} {2014})}\BibitemShut
  {NoStop}%
\bibitem [{\citenamefont {Fu}\ \emph {et~al.}(2017)\citenamefont {Fu},
  \citenamefont {Chang}, \citenamefont {Ye},\ and\ \citenamefont
  {Yin}}]{FU2017107}%
  \BibitemOpen
  \bibfield  {author} {\bibinfo {author} {\bibfnamefont {K.}~\bibnamefont
  {Fu}}, \bibinfo {author} {\bibfnamefont {L.}~\bibnamefont {Chang}}, \bibinfo
  {author} {\bibfnamefont {L.}~\bibnamefont {Ye}}, \ and\ \bibinfo {author}
  {\bibfnamefont {Y.}~\bibnamefont {Yin}},\ }\href {\doibase
  https://doi.org/10.1016/j.vacuum.2017.07.027} {\bibfield  {journal} {\bibinfo
   {journal} {Vacuum}\ }\textbf {\bibinfo {volume} {144}},\ \bibinfo {pages}
  {107 } (\bibinfo {year} {2017})}\BibitemShut {NoStop}%
\bibitem [{\citenamefont {Rasel}\ \emph {et~al.}(2011)\citenamefont {Rasel},
  \citenamefont {Wang}, \citenamefont {Ku}, \citenamefont {Byeon},
  \citenamefont {Kim},\ and\ \citenamefont {Song}}]{machanick1998design}%
  \BibitemOpen
  \bibfield  {author} {\bibinfo {author} {\bibfnamefont {S.}~\bibnamefont
  {Rasel}}, \bibinfo {author} {\bibfnamefont {Y.}~\bibnamefont {Wang}},
  \bibinfo {author} {\bibfnamefont {H.}~\bibnamefont {Ku}}, \bibinfo {author}
  {\bibfnamefont {J.}~\bibnamefont {Byeon}}, \bibinfo {author} {\bibfnamefont
  {T.}~\bibnamefont {Kim}}, \ and\ \bibinfo {author} {\bibfnamefont
  {J.}~\bibnamefont {Song}},\ }in\ \href@noop {} {\emph {\bibinfo {booktitle}
  {Proceedings of 18th international conference on composite materials}}}\
  (\bibinfo {year} {2011})\BibitemShut {NoStop}%
\bibitem [{\citenamefont {Li}\ and\ \citenamefont {Bhushan}(1998)}]{LI1998214}%
  \BibitemOpen
  \bibfield  {author} {\bibinfo {author} {\bibfnamefont {X.}~\bibnamefont
  {Li}}\ and\ \bibinfo {author} {\bibfnamefont {B.}~\bibnamefont {Bhushan}},\
  }\href {\doibase https://doi.org/10.1016/S0040-6090(97)00788-8} {\bibfield
  {journal} {\bibinfo  {journal} {Thin Solid Films}\ }\textbf {\bibinfo
  {volume} {315}},\ \bibinfo {pages} {214 } (\bibinfo {year}
  {1998})}\BibitemShut {NoStop}%
\bibitem [{\citenamefont {Schiffmann}(2011)}]{schiffmann2011determination}%
  \BibitemOpen
  \bibfield  {author} {\bibinfo {author} {\bibfnamefont {K.~I.}\ \bibnamefont
  {Schiffmann}},\ }\href@noop {} {\bibfield  {journal} {\bibinfo  {journal}
  {Phil. Mag.}\ }\textbf {\bibinfo {volume} {91}},\ \bibinfo {pages} {1163}
  (\bibinfo {year} {2011})}\BibitemShut {NoStop}%
\bibitem [{\citenamefont {Sinclair}\ and\ \citenamefont
  {Lawn}(1972)}]{sinclair_atomistic_1972}%
  \BibitemOpen
  \bibfield  {author} {\bibinfo {author} {\bibfnamefont {J.~E.}\ \bibnamefont
  {Sinclair}}\ and\ \bibinfo {author} {\bibfnamefont {B.~R.}\ \bibnamefont
  {Lawn}},\ }\href {\doibase 10.1098/rspa.1972.0102} {\bibfield  {journal}
  {\bibinfo  {journal} {Proc. R. Soc. Lond. A}\ }\textbf {\bibinfo {volume}
  {329}},\ \bibinfo {pages} {83} (\bibinfo {year} {1972})}\BibitemShut
  {NoStop}%
\bibitem [{\citenamefont {Bitzek}\ \emph {et~al.}(2015)\citenamefont {Bitzek},
  \citenamefont {Kermode},\ and\ \citenamefont
  {Gumbsch}}]{bitzek_atomistic_2015}%
  \BibitemOpen
  \bibfield  {author} {\bibinfo {author} {\bibfnamefont {E.}~\bibnamefont
  {Bitzek}}, \bibinfo {author} {\bibfnamefont {J.~R.}\ \bibnamefont {Kermode}},
  \ and\ \bibinfo {author} {\bibfnamefont {P.}~\bibnamefont {Gumbsch}},\ }\href
  {\doibase 10/f655cv} {\bibfield  {journal} {\bibinfo  {journal} {Int. J.
  Fract.}\ }\textbf {\bibinfo {volume} {191}},\ \bibinfo {pages} {13} (\bibinfo
  {year} {2015})}\BibitemShut {NoStop}%
\bibitem [{\citenamefont {Andric}\ and\ \citenamefont
  {Curtin}(2018)}]{andric_atomistic_2018}%
  \BibitemOpen
  \bibfield  {author} {\bibinfo {author} {\bibfnamefont {P.}~\bibnamefont
  {Andric}}\ and\ \bibinfo {author} {\bibfnamefont {W.~A.}\ \bibnamefont
  {Curtin}},\ }\href {\doibase 10.1088/1361-651X/aae40c} {\bibfield  {journal}
  {\bibinfo  {journal} {Modelling Simul. Mater. Sci. Eng.}\ }\textbf {\bibinfo
  {volume} {27}},\ \bibinfo {pages} {013001} (\bibinfo {year}
  {2018})}\BibitemShut {NoStop}%
\bibitem [{\citenamefont {Broughton}\ \emph {et~al.}(1999)\citenamefont
  {Broughton}, \citenamefont {Abraham}, \citenamefont {Bernstein},\ and\
  \citenamefont {Kaxiras}}]{broughton_concurrent_1999}%
  \BibitemOpen
  \bibfield  {author} {\bibinfo {author} {\bibfnamefont {J.~Q.}\ \bibnamefont
  {Broughton}}, \bibinfo {author} {\bibfnamefont {F.~F.}\ \bibnamefont
  {Abraham}}, \bibinfo {author} {\bibfnamefont {N.}~\bibnamefont {Bernstein}},
  \ and\ \bibinfo {author} {\bibfnamefont {E.}~\bibnamefont {Kaxiras}},\ }\href
  {\doibase 10/cm89fm} {\bibfield  {journal} {\bibinfo  {journal} {Phys. Rev.
  B}\ }\textbf {\bibinfo {volume} {60}},\ \bibinfo {pages} {2391} (\bibinfo
  {year} {1999})}\BibitemShut {NoStop}%
\bibitem [{\citenamefont {Abraham}\ \emph {et~al.}(2000)\citenamefont
  {Abraham}, \citenamefont {Bernstein}, \citenamefont {Broughton},\ and\
  \citenamefont {Hess}}]{abraham_dynamic_2000}%
  \BibitemOpen
  \bibfield  {author} {\bibinfo {author} {\bibfnamefont {F.~F.}\ \bibnamefont
  {Abraham}}, \bibinfo {author} {\bibfnamefont {N.}~\bibnamefont {Bernstein}},
  \bibinfo {author} {\bibfnamefont {J.~Q.}\ \bibnamefont {Broughton}}, \ and\
  \bibinfo {author} {\bibfnamefont {D.}~\bibnamefont {Hess}},\ }\href {\doibase
  10/d99rgb} {\bibfield  {journal} {\bibinfo  {journal} {MRS Bull.}\ }\textbf
  {\bibinfo {volume} {25}},\ \bibinfo {pages} {27} (\bibinfo {year}
  {2000})}\BibitemShut {NoStop}%
\bibitem [{\citenamefont {Pérez}\ and\ \citenamefont
  {Gumbsch}(2000{\natexlab{a}})}]{perez_directional_2000}%
  \BibitemOpen
  \bibfield  {author} {\bibinfo {author} {\bibfnamefont {R.}~\bibnamefont
  {Pérez}}\ and\ \bibinfo {author} {\bibfnamefont {P.}~\bibnamefont
  {Gumbsch}},\ }\href {\doibase 10.1103/PhysRevLett.84.5347} {\bibfield
  {journal} {\bibinfo  {journal} {Phys. Rev. Lett.}\ }\textbf {\bibinfo
  {volume} {84}},\ \bibinfo {pages} {5347} (\bibinfo {year}
  {2000}{\natexlab{a}})}\BibitemShut {NoStop}%
\bibitem [{\citenamefont {Pérez}\ and\ \citenamefont
  {Gumbsch}(2000{\natexlab{b}})}]{perez_ab_2000}%
  \BibitemOpen
  \bibfield  {author} {\bibinfo {author} {\bibfnamefont {R.}~\bibnamefont
  {Pérez}}\ and\ \bibinfo {author} {\bibfnamefont {P.}~\bibnamefont
  {Gumbsch}},\ }\href {\doibase 10.1016/S1359-6454(00)00238-X} {\bibfield
  {journal} {\bibinfo  {journal} {Acta Mater.}\ }\textbf {\bibinfo {volume}
  {48}},\ \bibinfo {pages} {4517} (\bibinfo {year}
  {2000}{\natexlab{b}})}\BibitemShut {NoStop}%
\bibitem [{\citenamefont {Bernstein}\ and\ \citenamefont
  {Hess}(2003)}]{bernstein_lattice_2003}%
  \BibitemOpen
  \bibfield  {author} {\bibinfo {author} {\bibfnamefont {N.}~\bibnamefont
  {Bernstein}}\ and\ \bibinfo {author} {\bibfnamefont {D.}~\bibnamefont
  {Hess}},\ }\href {\doibase 10/btcjw7} {\bibfield  {journal} {\bibinfo
  {journal} {Phys. Rev. Lett.}\ }\textbf {\bibinfo {volume} {91}},\ \bibinfo
  {pages} {025501} (\bibinfo {year} {2003})}\BibitemShut {NoStop}%
\bibitem [{\citenamefont {Zhu}\ \emph {et~al.}(2004)\citenamefont {Zhu},
  \citenamefont {Li},\ and\ \citenamefont {Yip}}]{zhu_atomistic_2004}%
  \BibitemOpen
  \bibfield  {author} {\bibinfo {author} {\bibfnamefont {T.}~\bibnamefont
  {Zhu}}, \bibinfo {author} {\bibfnamefont {J.}~\bibnamefont {Li}}, \ and\
  \bibinfo {author} {\bibfnamefont {S.}~\bibnamefont {Yip}},\ }\href {\doibase
  10/c9bm36} {\bibfield  {journal} {\bibinfo  {journal} {Phys. Rev. Lett.}\
  }\textbf {\bibinfo {volume} {93}},\ \bibinfo {pages} {205504} (\bibinfo
  {year} {2004})}\BibitemShut {NoStop}%
\bibitem [{\citenamefont {Zhu}\ \emph {et~al.}(2006)\citenamefont {Zhu},
  \citenamefont {Li},\ and\ \citenamefont {Yip}}]{zhu_atomistic_2006}%
  \BibitemOpen
  \bibfield  {author} {\bibinfo {author} {\bibfnamefont {T.}~\bibnamefont
  {Zhu}}, \bibinfo {author} {\bibfnamefont {J.}~\bibnamefont {Li}}, \ and\
  \bibinfo {author} {\bibfnamefont {S.}~\bibnamefont {Yip}},\ }\href {\doibase
  10.1098/rspa.2005.1567} {\bibfield  {journal} {\bibinfo  {journal} {Proc. R.
  Soc. A}\ }\textbf {\bibinfo {volume} {462}},\ \bibinfo {pages} {1741}
  (\bibinfo {year} {2006})}\BibitemShut {NoStop}%
\bibitem [{\citenamefont {Buehler}\ \emph {et~al.}(2006)\citenamefont
  {Buehler}, \citenamefont {van Duin},\ and\ \citenamefont
  {Goddard}}]{buehler_multiparadigm_2006}%
  \BibitemOpen
  \bibfield  {author} {\bibinfo {author} {\bibfnamefont {M.~J.}\ \bibnamefont
  {Buehler}}, \bibinfo {author} {\bibfnamefont {A.~C.~T.}\ \bibnamefont {van
  Duin}}, \ and\ \bibinfo {author} {\bibfnamefont {W.~A.}\ \bibnamefont
  {Goddard}},\ }\href {\doibase 10.1103/PhysRevLett.96.095505} {\bibfield
  {journal} {\bibinfo  {journal} {Phys. Rev. Lett.}\ }\textbf {\bibinfo
  {volume} {96}},\ \bibinfo {pages} {095505} (\bibinfo {year}
  {2006})}\BibitemShut {NoStop}%
\bibitem [{\citenamefont {Buehler}\ \emph {et~al.}(2007)\citenamefont
  {Buehler}, \citenamefont {Tang}, \citenamefont {van Duin},\ and\
  \citenamefont {Goddard}}]{buehler_threshold_2007}%
  \BibitemOpen
  \bibfield  {author} {\bibinfo {author} {\bibfnamefont {M.~J.}\ \bibnamefont
  {Buehler}}, \bibinfo {author} {\bibfnamefont {H.}~\bibnamefont {Tang}},
  \bibinfo {author} {\bibfnamefont {A.~C.~T.}\ \bibnamefont {van Duin}}, \ and\
  \bibinfo {author} {\bibfnamefont {W.~A.}\ \bibnamefont {Goddard},
  \bibfnamefont {III}},\ }\href@noop {} {\bibfield  {journal} {\bibinfo
  {journal} {Phys. Rev. Lett.}\ }\textbf {\bibinfo {volume} {99}},\ \bibinfo
  {pages} {165502} (\bibinfo {year} {2007})}\BibitemShut {NoStop}%
\bibitem [{\citenamefont {Kermode}\ \emph {et~al.}(2008)\citenamefont
  {Kermode}, \citenamefont {Albaret}, \citenamefont {Sherman}, \citenamefont
  {Bernstein}, \citenamefont {Gumbsch}, \citenamefont {Payne}, \citenamefont
  {Csányi},\ and\ \citenamefont {de~Vita}}]{kermode_low-speed_2008}%
  \BibitemOpen
  \bibfield  {author} {\bibinfo {author} {\bibfnamefont {J.~R.}\ \bibnamefont
  {Kermode}}, \bibinfo {author} {\bibfnamefont {T.}~\bibnamefont {Albaret}},
  \bibinfo {author} {\bibfnamefont {D.}~\bibnamefont {Sherman}}, \bibinfo
  {author} {\bibfnamefont {N.}~\bibnamefont {Bernstein}}, \bibinfo {author}
  {\bibfnamefont {P.}~\bibnamefont {Gumbsch}}, \bibinfo {author} {\bibfnamefont
  {M.~C.}\ \bibnamefont {Payne}}, \bibinfo {author} {\bibfnamefont
  {G.}~\bibnamefont {Csányi}}, \ and\ \bibinfo {author} {\bibfnamefont
  {A.}~\bibnamefont {de~Vita}},\ }\href {\doibase 10.1038/nature07297}
  {\bibfield  {journal} {\bibinfo  {journal} {Nature}\ }\textbf {\bibinfo
  {volume} {455}},\ \bibinfo {pages} {1224} (\bibinfo {year}
  {2008})}\BibitemShut {NoStop}%
\bibitem [{\citenamefont {Kermode}\ \emph {et~al.}(2013)\citenamefont
  {Kermode}, \citenamefont {Ben-Bashat}, \citenamefont {Atrash}, \citenamefont
  {Cilliers}, \citenamefont {Sherman},\ and\ \citenamefont
  {de~Vita}}]{kermode_macroscopic_2013}%
  \BibitemOpen
  \bibfield  {author} {\bibinfo {author} {\bibfnamefont {J.~R.}\ \bibnamefont
  {Kermode}}, \bibinfo {author} {\bibfnamefont {L.}~\bibnamefont {Ben-Bashat}},
  \bibinfo {author} {\bibfnamefont {F.}~\bibnamefont {Atrash}}, \bibinfo
  {author} {\bibfnamefont {J.~J.}\ \bibnamefont {Cilliers}}, \bibinfo {author}
  {\bibfnamefont {D.}~\bibnamefont {Sherman}}, \ and\ \bibinfo {author}
  {\bibfnamefont {A.}~\bibnamefont {de~Vita}},\ }\href {\doibase 10/gft56x}
  {\bibfield  {journal} {\bibinfo  {journal} {Nat. Comm.}\ }\textbf {\bibinfo
  {volume} {4}},\ \bibinfo {pages} {2441} (\bibinfo {year} {2013})}\BibitemShut
  {NoStop}%
\bibitem [{\citenamefont {Gleizer}\ \emph {et~al.}(2014)\citenamefont
  {Gleizer}, \citenamefont {Peralta}, \citenamefont {Kermode}, \citenamefont
  {De~Vita},\ and\ \citenamefont {Sherman}}]{gleizer_dissociative_2014}%
  \BibitemOpen
  \bibfield  {author} {\bibinfo {author} {\bibfnamefont {A.}~\bibnamefont
  {Gleizer}}, \bibinfo {author} {\bibfnamefont {G.}~\bibnamefont {Peralta}},
  \bibinfo {author} {\bibfnamefont {J.~R.}\ \bibnamefont {Kermode}}, \bibinfo
  {author} {\bibfnamefont {A.}~\bibnamefont {De~Vita}}, \ and\ \bibinfo
  {author} {\bibfnamefont {D.}~\bibnamefont {Sherman}},\ }\href {\doibase
  10/gft575} {\bibfield  {journal} {\bibinfo  {journal} {Phys. Rev. Lett.}\
  }\textbf {\bibinfo {volume} {112}},\ \bibinfo {pages} {115501} (\bibinfo
  {year} {2014})}\BibitemShut {NoStop}%
\bibitem [{\citenamefont {Kermode}\ \emph {et~al.}(2015)\citenamefont
  {Kermode}, \citenamefont {Gleizer}, \citenamefont {Kovel}, \citenamefont
  {Pastewka}, \citenamefont {Csányi}, \citenamefont {Sherman},\ and\
  \citenamefont {De~Vita}}]{kermode_low_2015}%
  \BibitemOpen
  \bibfield  {author} {\bibinfo {author} {\bibfnamefont {J.~R.}\ \bibnamefont
  {Kermode}}, \bibinfo {author} {\bibfnamefont {A.}~\bibnamefont {Gleizer}},
  \bibinfo {author} {\bibfnamefont {G.}~\bibnamefont {Kovel}}, \bibinfo
  {author} {\bibfnamefont {L.}~\bibnamefont {Pastewka}}, \bibinfo {author}
  {\bibfnamefont {G.}~\bibnamefont {Csányi}}, \bibinfo {author} {\bibfnamefont
  {D.}~\bibnamefont {Sherman}}, \ and\ \bibinfo {author} {\bibfnamefont
  {A.}~\bibnamefont {De~Vita}},\ }\href {\doibase
  10.1103/PhysRevLett.115.135501} {\bibfield  {journal} {\bibinfo  {journal}
  {Phys. Rev. Lett.}\ }\textbf {\bibinfo {volume} {115}},\ \bibinfo {pages}
  {135501} (\bibinfo {year} {2015})}\BibitemShut {NoStop}%
\bibitem [{\citenamefont {Pastewka}\ \emph {et~al.}(2008)\citenamefont
  {Pastewka}, \citenamefont {Pou}, \citenamefont {Pérez}, \citenamefont
  {Gumbsch},\ and\ \citenamefont {Moseler}}]{pastewka_describing_2008}%
  \BibitemOpen
  \bibfield  {author} {\bibinfo {author} {\bibfnamefont {L.}~\bibnamefont
  {Pastewka}}, \bibinfo {author} {\bibfnamefont {P.}~\bibnamefont {Pou}},
  \bibinfo {author} {\bibfnamefont {R.}~\bibnamefont {Pérez}}, \bibinfo
  {author} {\bibfnamefont {P.}~\bibnamefont {Gumbsch}}, \ and\ \bibinfo
  {author} {\bibfnamefont {M.}~\bibnamefont {Moseler}},\ }\href {\doibase
  10.1103/PhysRevB.78.161402} {\bibfield  {journal} {\bibinfo  {journal} {Phys.
  Rev. B}\ }\textbf {\bibinfo {volume} {78}},\ \bibinfo {pages} {161402(R)}
  (\bibinfo {year} {2008})}\BibitemShut {NoStop}%
\bibitem [{\citenamefont {Pastewka}\ \emph {et~al.}(2013)\citenamefont
  {Pastewka}, \citenamefont {Klemenz}, \citenamefont {Gumbsch},\ and\
  \citenamefont {Moseler}}]{pastewka_screened_2013}%
  \BibitemOpen
  \bibfield  {author} {\bibinfo {author} {\bibfnamefont {L.}~\bibnamefont
  {Pastewka}}, \bibinfo {author} {\bibfnamefont {A.}~\bibnamefont {Klemenz}},
  \bibinfo {author} {\bibfnamefont {P.}~\bibnamefont {Gumbsch}}, \ and\
  \bibinfo {author} {\bibfnamefont {M.}~\bibnamefont {Moseler}},\ }\href
  {\doibase 10/gft5th} {\bibfield  {journal} {\bibinfo  {journal} {Phys. Rev.
  B}\ }\textbf {\bibinfo {volume} {87}},\ \bibinfo {pages} {205410} (\bibinfo
  {year} {2013})}\BibitemShut {NoStop}%
\bibitem [{\citenamefont {Falk}(1999)}]{falk_molecular-dynamics_1999}%
  \BibitemOpen
  \bibfield  {author} {\bibinfo {author} {\bibfnamefont {M.~L.}\ \bibnamefont
  {Falk}},\ }\href {\doibase 10.1103/PhysRevB.60.7062} {\bibfield  {journal}
  {\bibinfo  {journal} {Phys. Rev. B}\ }\textbf {\bibinfo {volume} {60}},\
  \bibinfo {pages} {7062} (\bibinfo {year} {1999})}\BibitemShut {NoStop}%
\bibitem [{\citenamefont {Falk}\ and\ \citenamefont
  {Langer}(2000)}]{falk_simulation_2000}%
  \BibitemOpen
  \bibfield  {author} {\bibinfo {author} {\bibfnamefont {M.~L.}\ \bibnamefont
  {Falk}}\ and\ \bibinfo {author} {\bibfnamefont {J.~S.}\ \bibnamefont
  {Langer}},\ }\href {\doibase 10/c36mzc} {\bibfield  {journal} {\bibinfo
  {journal} {MRS Bull.}\ }\textbf {\bibinfo {volume} {25}},\ \bibinfo {pages}
  {40} (\bibinfo {year} {2000})}\BibitemShut {NoStop}%
\bibitem [{\citenamefont {Khosrownejad}\ and\ \citenamefont
  {Curtin}(2017)}]{Khosrownejad_2017}%
  \BibitemOpen
  \bibfield  {author} {\bibinfo {author} {\bibfnamefont {S.~M.}\ \bibnamefont
  {Khosrownejad}}\ and\ \bibinfo {author} {\bibfnamefont {W.~A.}\ \bibnamefont
  {Curtin}},\ }\href {\doibase 10.1016/j.jmps.2017.06.010} {\bibfield
  {journal} {\bibinfo  {journal} {J. Mech. Phys. Solids}\ }\textbf {\bibinfo
  {volume} {107}},\ \bibinfo {pages} {18. 542} (\bibinfo {year}
  {2017})}\BibitemShut {NoStop}%
\bibitem [{\citenamefont {Fyta}\ \emph {et~al.}(2006)\citenamefont {Fyta},
  \citenamefont {Remediakis}, \citenamefont {Kelires},\ and\ \citenamefont
  {Papaconstantopoulos}}]{fyta2006insights}%
  \BibitemOpen
  \bibfield  {author} {\bibinfo {author} {\bibfnamefont {M.~G.}\ \bibnamefont
  {Fyta}}, \bibinfo {author} {\bibfnamefont {I.~N.}\ \bibnamefont
  {Remediakis}}, \bibinfo {author} {\bibfnamefont {P.~C.}\ \bibnamefont
  {Kelires}}, \ and\ \bibinfo {author} {\bibfnamefont {D.~A.}\ \bibnamefont
  {Papaconstantopoulos}},\ }\href@noop {} {\bibfield  {journal} {\bibinfo
  {journal} {Phys. Rev. Lett.}\ }\textbf {\bibinfo {volume} {96}},\ \bibinfo
  {pages} {185503} (\bibinfo {year} {2006})}\BibitemShut {NoStop}%
\bibitem [{\citenamefont {Lee}\ \emph {et~al.}(2018)\citenamefont {Lee},
  \citenamefont {Chung}, \citenamefont {Na},\ and\ \citenamefont
  {Beom}}]{lee2018atomistic}%
  \BibitemOpen
  \bibfield  {author} {\bibinfo {author} {\bibfnamefont {G.~H.}\ \bibnamefont
  {Lee}}, \bibinfo {author} {\bibfnamefont {Y.~J.}\ \bibnamefont {Chung}},
  \bibinfo {author} {\bibfnamefont {S.~M.}\ \bibnamefont {Na}}, \ and\ \bibinfo
  {author} {\bibfnamefont {H.~G.}\ \bibnamefont {Beom}},\ }\href@noop {}
  {\bibfield  {journal} {\bibinfo  {journal} {J. Mech. Sci. Technol.}\ }\textbf
  {\bibinfo {volume} {32}},\ \bibinfo {pages} {3765} (\bibinfo {year}
  {2018})}\BibitemShut {NoStop}%
\bibitem [{\citenamefont {Sedighiani}\ \emph {et~al.}(2011)\citenamefont
  {Sedighiani}, \citenamefont {Mosayebnejad}, \citenamefont {Ehsasi},\ and\
  \citenamefont {Sahraei}}]{sedighiani2011effect}%
  \BibitemOpen
  \bibfield  {author} {\bibinfo {author} {\bibfnamefont {K.}~\bibnamefont
  {Sedighiani}}, \bibinfo {author} {\bibfnamefont {J.}~\bibnamefont
  {Mosayebnejad}}, \bibinfo {author} {\bibfnamefont {H.}~\bibnamefont
  {Ehsasi}}, \ and\ \bibinfo {author} {\bibfnamefont {H.}~\bibnamefont
  {Sahraei}},\ }\href@noop {} {\bibfield  {journal} {\bibinfo  {journal}
  {Procedia Engineer.}\ }\textbf {\bibinfo {volume} {10}},\ \bibinfo {pages}
  {774} (\bibinfo {year} {2011})}\BibitemShut {NoStop}%
\bibitem [{\citenamefont {Sinclair}(1975)}]{sinclair1975influence}%
  \BibitemOpen
  \bibfield  {author} {\bibinfo {author} {\bibfnamefont {J.}~\bibnamefont
  {Sinclair}},\ }\href@noop {} {\bibfield  {journal} {\bibinfo  {journal}
  {Phil. Mag.}\ }\textbf {\bibinfo {volume} {31}},\ \bibinfo {pages} {647}
  (\bibinfo {year} {1975})}\BibitemShut {NoStop}%
\bibitem [{\citenamefont {Buze}\ and\ \citenamefont
  {Kermode}(2020)}]{buze_numerical-continuation-enhanced_2020}%
  \BibitemOpen
  \bibfield  {author} {\bibinfo {author} {\bibfnamefont {M.}~\bibnamefont
  {Buze}}\ and\ \bibinfo {author} {\bibfnamefont {J.~R.}\ \bibnamefont
  {Kermode}},\ }\href@noop {} {\bibfield  {journal} {\bibinfo  {journal}
  {arXiv:2008.12822}\ } (\bibinfo {year} {2020})}\BibitemShut {NoStop}%
\bibitem [{\citenamefont {Jana}\ \emph {et~al.}(2019)\citenamefont {Jana},
  \citenamefont {Savio}, \citenamefont {Deringer},\ and\ \citenamefont
  {Pastewka}}]{Jana_2019}%
  \BibitemOpen
  \bibfield  {author} {\bibinfo {author} {\bibfnamefont {R.}~\bibnamefont
  {Jana}}, \bibinfo {author} {\bibfnamefont {D.}~\bibnamefont {Savio}},
  \bibinfo {author} {\bibfnamefont {V.~L.}\ \bibnamefont {Deringer}}, \ and\
  \bibinfo {author} {\bibfnamefont {L.}~\bibnamefont {Pastewka}},\ }\href
  {\doibase 10.1088/1361-651x/ab45da} {\bibfield  {journal} {\bibinfo
  {journal} {Modelling Simul. Mater. Sci. Eng.}\ }\textbf {\bibinfo {volume}
  {27}},\ \bibinfo {pages} {085009} (\bibinfo {year} {2019})}\BibitemShut
  {NoStop}%
\bibitem [{\citenamefont {Jana}\ \emph {et~al.}(2020)\citenamefont {Jana},
  \citenamefont {Lautz}, \citenamefont {Khosrownejad}, \citenamefont {Andrews},
  \citenamefont {Moseler},\ and\ \citenamefont {Pastewka}}]{Jana_2020}%
  \BibitemOpen
  \bibfield  {author} {\bibinfo {author} {\bibfnamefont {R.}~\bibnamefont
  {Jana}}, \bibinfo {author} {\bibfnamefont {J.~v.}\ \bibnamefont {Lautz}},
  \bibinfo {author} {\bibfnamefont {S.~M.}\ \bibnamefont {Khosrownejad}},
  \bibinfo {author} {\bibfnamefont {W.~B.}\ \bibnamefont {Andrews}}, \bibinfo
  {author} {\bibfnamefont {M.}~\bibnamefont {Moseler}}, \ and\ \bibinfo
  {author} {\bibfnamefont {L.}~\bibnamefont {Pastewka}},\ }\href {\doibase
  10.1088/2515-7639/ab953c} {\bibfield  {journal} {\bibinfo  {journal} {J.
  Phys. Mater.}\ }\textbf {\bibinfo {volume} {3}},\ \bibinfo {pages} {035005}
  (\bibinfo {year} {2020})}\BibitemShut {NoStop}%
\bibitem [{\citenamefont {Drucker}\ and\ \citenamefont
  {Prager}(1952)}]{drucker_soil_1952}%
  \BibitemOpen
  \bibfield  {author} {\bibinfo {author} {\bibfnamefont {D.~C.}\ \bibnamefont
  {Drucker}}\ and\ \bibinfo {author} {\bibfnamefont {W.}~\bibnamefont
  {Prager}},\ }\href {\doibase 10/gft5tj} {\bibfield  {journal} {\bibinfo
  {journal} {Q. Appl. Math.}\ }\textbf {\bibinfo {volume} {10}},\ \bibinfo
  {pages} {157} (\bibinfo {year} {1952})}\BibitemShut {NoStop}%
\bibitem [{\citenamefont {Drugan}\ \emph {et~al.}(1982)\citenamefont {Drugan},
  \citenamefont {Rice},\ and\ \citenamefont {Sham}}]{DRUGAN1982447}%
  \BibitemOpen
  \bibfield  {author} {\bibinfo {author} {\bibfnamefont {W.}~\bibnamefont
  {Drugan}}, \bibinfo {author} {\bibfnamefont {J.}~\bibnamefont {Rice}}, \ and\
  \bibinfo {author} {\bibfnamefont {T.-L.}\ \bibnamefont {Sham}},\ }\href
  {\doibase https://doi.org/10.1016/0022-5096(82)90027-8} {\bibfield  {journal}
  {\bibinfo  {journal} {J. Mech. Phys. Solids}\ }\textbf {\bibinfo {volume}
  {30}},\ \bibinfo {pages} {447 } (\bibinfo {year} {1982})}\BibitemShut
  {NoStop}%
\bibitem [{\citenamefont {Tersoff}(1989)}]{tersoff_modeling_1989}%
  \BibitemOpen
  \bibfield  {author} {\bibinfo {author} {\bibfnamefont {J.}~\bibnamefont
  {Tersoff}},\ }\href {\doibase 10.1103/PhysRevB.39.5566} {\bibfield  {journal}
  {\bibinfo  {journal} {Phys. Rev. B}\ }\textbf {\bibinfo {volume} {39}},\
  \bibinfo {pages} {5566} (\bibinfo {year} {1989})}\BibitemShut {NoStop}%
\bibitem [{\citenamefont {McCulloch}\ \emph
  {et~al.}(2000{\natexlab{a}})\citenamefont {McCulloch}, \citenamefont
  {McKenzie},\ and\ \citenamefont {Goringe}}]{mcculloch2000ab}%
  \BibitemOpen
  \bibfield  {author} {\bibinfo {author} {\bibfnamefont {D.}~\bibnamefont
  {McCulloch}}, \bibinfo {author} {\bibfnamefont {D.}~\bibnamefont {McKenzie}},
  \ and\ \bibinfo {author} {\bibfnamefont {C.}~\bibnamefont {Goringe}},\
  }\href@noop {} {\bibfield  {journal} {\bibinfo  {journal} {Phys. Rev. B}\
  }\textbf {\bibinfo {volume} {61}},\ \bibinfo {pages} {2349} (\bibinfo {year}
  {2000}{\natexlab{a}})}\BibitemShut {NoStop}%
\bibitem [{\citenamefont {McCulloch}\ \emph
  {et~al.}(2000{\natexlab{b}})\citenamefont {McCulloch}, \citenamefont
  {McKenzie},\ and\ \citenamefont {Goringe}}]{mcculloch_ab_2000}%
  \BibitemOpen
  \bibfield  {author} {\bibinfo {author} {\bibfnamefont {D.~G.}\ \bibnamefont
  {McCulloch}}, \bibinfo {author} {\bibfnamefont {D.~R.}\ \bibnamefont
  {McKenzie}}, \ and\ \bibinfo {author} {\bibfnamefont {C.~M.}\ \bibnamefont
  {Goringe}},\ }\href {\doibase 10.1103/PhysRevB.61.2349} {\bibfield  {journal}
  {\bibinfo  {journal} {Phys. Rev. B}\ }\textbf {\bibinfo {volume} {61}},\
  \bibinfo {pages} {2349} (\bibinfo {year} {2000}{\natexlab{b}})}\BibitemShut
  {NoStop}%
\bibitem [{\citenamefont {de~Tomas}\ \emph {et~al.}(2016)\citenamefont
  {de~Tomas}, \citenamefont {Suarez-Martinez},\ and\ \citenamefont
  {Marks}}]{de_tomas_graphitization_2016}%
  \BibitemOpen
  \bibfield  {author} {\bibinfo {author} {\bibfnamefont {C.}~\bibnamefont
  {de~Tomas}}, \bibinfo {author} {\bibfnamefont {I.}~\bibnamefont
  {Suarez-Martinez}}, \ and\ \bibinfo {author} {\bibfnamefont {N.~A.}\
  \bibnamefont {Marks}},\ }\href {\doibase 10.1016/j.carbon.2016.08.024}
  {\bibfield  {journal} {\bibinfo  {journal} {Carbon}\ }\textbf {\bibinfo
  {volume} {109}},\ \bibinfo {pages} {681} (\bibinfo {year}
  {2016})}\BibitemShut {NoStop}%
\bibitem [{\citenamefont {de~Tomas}\ \emph {et~al.}(2019)\citenamefont
  {de~Tomas}, \citenamefont {Aghajamali}, \citenamefont {Jones}, \citenamefont
  {Lim}, \citenamefont {Lopez}, \citenamefont {Suarez-Martinez},\ and\
  \citenamefont {Marks}}]{de_tomas_transferability_2019}%
  \BibitemOpen
  \bibfield  {author} {\bibinfo {author} {\bibfnamefont {C.}~\bibnamefont
  {de~Tomas}}, \bibinfo {author} {\bibfnamefont {A.}~\bibnamefont
  {Aghajamali}}, \bibinfo {author} {\bibfnamefont {J.~L.}\ \bibnamefont
  {Jones}}, \bibinfo {author} {\bibfnamefont {D.~J.}\ \bibnamefont {Lim}},
  \bibinfo {author} {\bibfnamefont {M.~J.}\ \bibnamefont {Lopez}}, \bibinfo
  {author} {\bibfnamefont {I.}~\bibnamefont {Suarez-Martinez}}, \ and\ \bibinfo
  {author} {\bibfnamefont {N.~A.}\ \bibnamefont {Marks}},\ }\href {\doibase
  10.1016/j.carbon.2019.07.074} {\bibfield  {journal} {\bibinfo  {journal}
  {Carbon}\ }\textbf {\bibinfo {volume} {155}},\ \bibinfo {pages} {624}
  (\bibinfo {year} {2019})}\BibitemShut {NoStop}%
\bibitem [{\citenamefont {Jäger}\ and\ \citenamefont
  {Albe}(2000)}]{jager_molecular-dynamics_2000}%
  \BibitemOpen
  \bibfield  {author} {\bibinfo {author} {\bibfnamefont {H.~U.}\ \bibnamefont
  {Jäger}}\ and\ \bibinfo {author} {\bibfnamefont {K.}~\bibnamefont {Albe}},\
  }\href {\doibase 10.1063/1.373787} {\bibfield  {journal} {\bibinfo  {journal}
  {J. Appl. Phys.}\ }\textbf {\bibinfo {volume} {88}},\ \bibinfo {pages} {1129}
  (\bibinfo {year} {2000})}\BibitemShut {NoStop}%
\bibitem [{\citenamefont {Belov}\ and\ \citenamefont
  {Jäger}(2002)}]{belov_calculation_2002}%
  \BibitemOpen
  \bibfield  {author} {\bibinfo {author} {\bibfnamefont {A.~Y.}\ \bibnamefont
  {Belov}}\ and\ \bibinfo {author} {\bibfnamefont {H.~U.}\ \bibnamefont
  {Jäger}},\ }\href@noop {} {\bibfield  {journal} {\bibinfo  {journal}
  {Comput. Mater. Sci.}\ }\textbf {\bibinfo {volume} {24}},\ \bibinfo {pages}
  {154} (\bibinfo {year} {2002})}\BibitemShut {NoStop}%
\bibitem [{\citenamefont {Jäger}\ and\ \citenamefont
  {Belov}(2003)}]{jager_ta-c_2003}%
  \BibitemOpen
  \bibfield  {author} {\bibinfo {author} {\bibfnamefont {H.~U.}\ \bibnamefont
  {Jäger}}\ and\ \bibinfo {author} {\bibfnamefont {A.~Y.}\ \bibnamefont
  {Belov}},\ }\href {\doibase 10.1103/PhysRevB.68.024201} {\bibfield  {journal}
  {\bibinfo  {journal} {Phys. Rev. B}\ }\textbf {\bibinfo {volume} {68}},\
  \bibinfo {pages} {24201} (\bibinfo {year} {2003})}\BibitemShut {NoStop}%
\bibitem [{\citenamefont {Moseler}\ \emph {et~al.}(2005)\citenamefont
  {Moseler}, \citenamefont {Gumbsch}, \citenamefont {Casiraghi}, \citenamefont
  {Ferrari},\ and\ \citenamefont {Robertson}}]{moseler_ultrasmoothness_2005}%
  \BibitemOpen
  \bibfield  {author} {\bibinfo {author} {\bibfnamefont {M.}~\bibnamefont
  {Moseler}}, \bibinfo {author} {\bibfnamefont {P.}~\bibnamefont {Gumbsch}},
  \bibinfo {author} {\bibfnamefont {C.}~\bibnamefont {Casiraghi}}, \bibinfo
  {author} {\bibfnamefont {A.~C.}\ \bibnamefont {Ferrari}}, \ and\ \bibinfo
  {author} {\bibfnamefont {J.}~\bibnamefont {Robertson}},\ }\href {\doibase
  10.1126/science.1114577} {\bibfield  {journal} {\bibinfo  {journal}
  {Science}\ }\textbf {\bibinfo {volume} {309}},\ \bibinfo {pages} {1545}
  (\bibinfo {year} {2005})}\BibitemShut {NoStop}%
\bibitem [{\citenamefont {Caro}\ \emph {et~al.}(2018)\citenamefont {Caro},
  \citenamefont {Deringer}, \citenamefont {Koskinen}, \citenamefont {Laurila},\
  and\ \citenamefont {Csányi}}]{caro_growth_2018}%
  \BibitemOpen
  \bibfield  {author} {\bibinfo {author} {\bibfnamefont {M.~A.}\ \bibnamefont
  {Caro}}, \bibinfo {author} {\bibfnamefont {V.~L.}\ \bibnamefont {Deringer}},
  \bibinfo {author} {\bibfnamefont {J.}~\bibnamefont {Koskinen}}, \bibinfo
  {author} {\bibfnamefont {T.}~\bibnamefont {Laurila}}, \ and\ \bibinfo
  {author} {\bibfnamefont {G.}~\bibnamefont {Csányi}},\ }\href {\doibase
  10/gc96hp} {\bibfield  {journal} {\bibinfo  {journal} {Phys. Rev. Lett.}\
  }\textbf {\bibinfo {volume} {120}},\ \bibinfo {pages} {166101} (\bibinfo
  {year} {2018})}\BibitemShut {NoStop}%
\bibitem [{\citenamefont {Hutchinson}(1968)}]{Hutchinson1968}%
  \BibitemOpen
  \bibfield  {author} {\bibinfo {author} {\bibfnamefont {J.~W.}\ \bibnamefont
  {Hutchinson}},\ }\href {\doibase 10.1016/0022-5096(68)90014-8} {\bibfield
  {journal} {\bibinfo  {journal} {J. Mech. Phys. Solids}\ }\textbf {\bibinfo
  {volume} {16}},\ \bibinfo {pages} {13} (\bibinfo {year} {1968})}\BibitemShut
  {NoStop}%
\bibitem [{\citenamefont {Rice}\ and\ \citenamefont
  {Rosengren}(1968)}]{rice_plane_1968}%
  \BibitemOpen
  \bibfield  {author} {\bibinfo {author} {\bibfnamefont {J.~R.}\ \bibnamefont
  {Rice}}\ and\ \bibinfo {author} {\bibfnamefont {G.~F.}\ \bibnamefont
  {Rosengren}},\ }\href {\doibase 10.1016/0022-5096(68)90013-6} {\bibfield
  {journal} {\bibinfo  {journal} {J. Mech. Phys. Solids}\ }\textbf {\bibinfo
  {volume} {16}},\ \bibinfo {pages} {1} (\bibinfo {year} {1968})}\BibitemShut
  {NoStop}%
\bibitem [{\citenamefont {Ramberg}\ and\ \citenamefont
  {Osgood}(1943)}]{ramberg_description_1943}%
  \BibitemOpen
  \bibfield  {author} {\bibinfo {author} {\bibfnamefont {W.}~\bibnamefont
  {Ramberg}}\ and\ \bibinfo {author} {\bibfnamefont {W.~R.}\ \bibnamefont
  {Osgood}},\ }\href
  {https://ntrs.nasa.gov/archive/nasa/casi.ntrs.nasa.gov/19930081614.pdf}
  {\emph {\bibinfo {title} {Description of stress–strain curves by three
  parameters}}},\ \bibinfo {type} {Tech. Rep.}\ \bibinfo {number} {902}\
  (\bibinfo  {institution} {National Advisory Committee for Aeronautics},\
  \bibinfo {address} {Washington, DC},\ \bibinfo {year} {1943})\BibitemShut
  {NoStop}%
\bibitem [{\citenamefont {{McMeeking}}(1977)}]{mcmeeking_finite_1977}%
  \BibitemOpen
  \bibfield  {author} {\bibinfo {author} {\bibfnamefont {R.~M.}\ \bibnamefont
  {{McMeeking}}},\ }\href {\doibase 10.1016/0022-5096(77)90003-5} {\bibfield
  {journal} {\bibinfo  {journal} {J. Mech. Phys. Solids}\ }\textbf {\bibinfo
  {volume} {25}},\ \bibinfo {pages} {357} (\bibinfo {year} {1977})}\BibitemShut
  {NoStop}%
\bibitem [{\citenamefont {Singh}\ \emph {et~al.}(2014)\citenamefont {Singh},
  \citenamefont {Kermode}, \citenamefont {De~Vita},\ and\ \citenamefont
  {Zimmerman}}]{singh_validity_2014}%
  \BibitemOpen
  \bibfield  {author} {\bibinfo {author} {\bibfnamefont {G.}~\bibnamefont
  {Singh}}, \bibinfo {author} {\bibfnamefont {J.~R.}\ \bibnamefont {Kermode}},
  \bibinfo {author} {\bibfnamefont {A.}~\bibnamefont {De~Vita}}, \ and\
  \bibinfo {author} {\bibfnamefont {R.~W.}\ \bibnamefont {Zimmerman}},\
  }\href@noop {} {\bibfield  {journal} {\bibinfo  {journal} {Int. J. Fract.}\
  }\textbf {\bibinfo {volume} {189}},\ \bibinfo {pages} {103} (\bibinfo {year}
  {2014})}\BibitemShut {NoStop}%
\bibitem [{\citenamefont {Bower}(2009)}]{bower2009applied}%
  \BibitemOpen
  \bibfield  {author} {\bibinfo {author} {\bibfnamefont {A.~F.}\ \bibnamefont
  {Bower}},\ }\href@noop {} {\emph {\bibinfo {title} {Applied Mechanics of
  Solids}}}\ (\bibinfo  {publisher} {CRC press},\ \bibinfo {year}
  {2009})\BibitemShut {NoStop}%
\bibitem [{\citenamefont {Banks}\ and\ \citenamefont
  {Garlick}(1984)}]{banks1984form}%
  \BibitemOpen
  \bibfield  {author} {\bibinfo {author} {\bibfnamefont {T.}~\bibnamefont
  {Banks}}\ and\ \bibinfo {author} {\bibfnamefont {A.}~\bibnamefont
  {Garlick}},\ }\href@noop {} {\bibfield  {journal} {\bibinfo  {journal} {Eng.
  Fract. Mech.}\ }\textbf {\bibinfo {volume} {19}},\ \bibinfo {pages} {571}
  (\bibinfo {year} {1984})}\BibitemShut {NoStop}%
\bibitem [{\citenamefont {Rice}(1974)}]{Rice_1974_17}%
  \BibitemOpen
  \bibfield  {author} {\bibinfo {author} {\bibfnamefont {J.}~\bibnamefont
  {Rice}},\ }\href {\doibase https://doi.org/10.1016/0022-5096(74)90010-6}
  {\bibfield  {journal} {\bibinfo  {journal} {J. Mech. Phys. Solids}\ }\textbf
  {\bibinfo {volume} {22}},\ \bibinfo {pages} {17 } (\bibinfo {year}
  {1974})}\BibitemShut {NoStop}%
\bibitem [{\citenamefont {Tvergaard}\ and\ \citenamefont
  {Hutchinson}(1992{\natexlab{a}})}]{tvergaard1992relation}%
  \BibitemOpen
  \bibfield  {author} {\bibinfo {author} {\bibfnamefont {V.}~\bibnamefont
  {Tvergaard}}\ and\ \bibinfo {author} {\bibfnamefont {J.~W.}\ \bibnamefont
  {Hutchinson}},\ }\href@noop {} {\bibfield  {journal} {\bibinfo  {journal} {J.
  Mech. Phys. Solids}\ }\textbf {\bibinfo {volume} {40}},\ \bibinfo {pages}
  {1377} (\bibinfo {year} {1992}{\natexlab{a}})}\BibitemShut {NoStop}%
\bibitem [{\citenamefont {Gupta}\ \emph {et~al.}(2014)\citenamefont {Gupta},
  \citenamefont {Alderliesten},\ and\ \citenamefont
  {Benedictus}}]{Gupta2014review}%
  \BibitemOpen
  \bibfield  {author} {\bibinfo {author} {\bibfnamefont {M.}~\bibnamefont
  {Gupta}}, \bibinfo {author} {\bibfnamefont {R.}~\bibnamefont {Alderliesten}},
  \ and\ \bibinfo {author} {\bibfnamefont {R.}~\bibnamefont {Benedictus}},\
  }\href@noop {} {\bibfield  {journal} {\bibinfo  {journal} {Eng. Fract.
  Mech.}\ }\textbf {\bibinfo {volume} {134}} (\bibinfo {year}
  {2014})}\BibitemShut {NoStop}%
\bibitem [{\citenamefont {Tvergaard}\ and\ \citenamefont
  {Hutchinson}(1992{\natexlab{b}})}]{TVERGAARD19921377}%
  \BibitemOpen
  \bibfield  {author} {\bibinfo {author} {\bibfnamefont {V.}~\bibnamefont
  {Tvergaard}}\ and\ \bibinfo {author} {\bibfnamefont {J.~W.}\ \bibnamefont
  {Hutchinson}},\ }\href {\doibase
  https://doi.org/10.1016/0022-5096(92)90020-3} {\bibfield  {journal} {\bibinfo
   {journal} {J. Mech. Phys. Solids}\ }\textbf {\bibinfo {volume} {40}},\
  \bibinfo {pages} {1377 } (\bibinfo {year} {1992}{\natexlab{b}})}\BibitemShut
  {NoStop}%
\bibitem [{\citenamefont {Xia}\ and\ \citenamefont {Shih}(1995)}]{XIA1995233}%
  \BibitemOpen
  \bibfield  {author} {\bibinfo {author} {\bibfnamefont {L.}~\bibnamefont
  {Xia}}\ and\ \bibinfo {author} {\bibfnamefont {C.}~\bibnamefont {Shih}},\
  }\href {\doibase https://doi.org/10.1016/0022-5096(94)00064-C} {\bibfield
  {journal} {\bibinfo  {journal} {J. Mech. Phys. Solids}\ }\textbf {\bibinfo
  {volume} {43}},\ \bibinfo {pages} {233 } (\bibinfo {year}
  {1995})}\BibitemShut {NoStop}%
\bibitem [{\citenamefont {Schaufler}\ \emph {et~al.}(2012)\citenamefont
  {Schaufler}, \citenamefont {Schmid}, \citenamefont {Durst},\ and\
  \citenamefont {Göken}}]{SCHAUFLER2012480}%
  \BibitemOpen
  \bibfield  {author} {\bibinfo {author} {\bibfnamefont {J.}~\bibnamefont
  {Schaufler}}, \bibinfo {author} {\bibfnamefont {C.}~\bibnamefont {Schmid}},
  \bibinfo {author} {\bibfnamefont {K.}~\bibnamefont {Durst}}, \ and\ \bibinfo
  {author} {\bibfnamefont {M.}~\bibnamefont {Göken}},\ }\href {\doibase
  https://doi.org/10.1016/j.tsf.2012.08.031} {\bibfield  {journal} {\bibinfo
  {journal} {Thin Solid Films}\ }\textbf {\bibinfo {volume} {522}},\ \bibinfo
  {pages} {480 } (\bibinfo {year} {2012})}\BibitemShut {NoStop}%
\bibitem [{\citenamefont {Ritchie}\ and\ \citenamefont
  {Thompson}(1985)}]{Ritchie1985}%
  \BibitemOpen
  \bibfield  {author} {\bibinfo {author} {\bibfnamefont {R.~O.}\ \bibnamefont
  {Ritchie}}\ and\ \bibinfo {author} {\bibfnamefont {A.~W.}\ \bibnamefont
  {Thompson}},\ }\href {\doibase 10.1007/BF02816050} {\bibfield  {journal}
  {\bibinfo  {journal} {Metall. Trans. A}\ }\textbf {\bibinfo {volume} {16}},\
  \bibinfo {pages} {233} (\bibinfo {year} {1985})}\BibitemShut {NoStop}%
\bibitem [{\citenamefont {Plimpton}(1995)}]{plimpton_fast_1995}%
  \BibitemOpen
  \bibfield  {author} {\bibinfo {author} {\bibfnamefont {S.}~\bibnamefont
  {Plimpton}},\ }\href {\doibase 10/cw7cjf} {\bibfield  {journal} {\bibinfo
  {journal} {J. Comput. Phys.}\ }\textbf {\bibinfo {volume} {117}},\ \bibinfo
  {pages} {1} (\bibinfo {year} {1995})}\BibitemShut {NoStop}%
\bibitem [{\citenamefont {Stukowski}(2010)}]{stukowski_visualization_2010}%
  \BibitemOpen
  \bibfield  {author} {\bibinfo {author} {\bibfnamefont {A.}~\bibnamefont
  {Stukowski}},\ }\href {\doibase 10/c2hx6n} {\bibfield  {journal} {\bibinfo
  {journal} {Modelling Simul. Mater. Sci. Eng.}\ }\textbf {\bibinfo {volume}
  {18}},\ \bibinfo {pages} {15012} (\bibinfo {year} {2010})}\BibitemShut
  {NoStop}%
\end{thebibliography}
